\documentstyle[12pt]{article}

\begin{document}

\baselineskip 6mm
\renewcommand{\thefootnote}{\fnsymbol{footnote}}

\newcommand{\nc}{\newcommand}
\newcommand{\rnc}{\renewcommand}


\rnc{\baselinestretch}{1.44}    
\setlength{\jot}{7pt}       
\rnc{\arraystretch}{1.64}   



\nc{\be}{\begin{equation}}

\nc{\ee}{\end{equation}}

\nc{\bea}{\begin{eqnarray}}

\nc{\eea}{\end{eqnarray}}

\nc{\ben}{\begin{eqnarray*}}

\nc{\een}{\end{eqnarray*}}

\nc{\xx}{\nonumber\\}

\nc{\ct}{\cite}

\nc{\la}{\label}

\nc{\ra}{\rightarrow}

\nc{\eq}[1]{(\ref{#1})}

\nc{\newcaption}[1]{\centerline{\parbox{6in}{\caption{#1}}}}

\newcommand{\sectiono}[1]{\section{#1}\setcounter{equation}{0}}
\renewcommand{\theequation}{\thesection.\arabic{equation}}

\nc{\fig}[3]{

\begin{figure}
\centerline{\epsfxsize=#1\epsfbox{#2.eps}}
\newcaption{#3. \label{#2}}
\end{figure}
}


\def\IR{{\hbox{{\rm I}\kern-.2em\hbox{\rm R}}}}
\def\IB{{\hbox{{\rm I}\kern-.2em\hbox{\rm B}}}}
\def\IN{{\hbox{{\rm I}\kern-.2em\hbox{\rm N}}}}
\def\IC{\,\,{\hbox{{\rm I}\kern-.59em\hbox{\bf C}}}}
\def\IZ{{\hbox{{\rm Z}\kern-.4em\hbox{\rm Z}}}}
\def\IP{{\hbox{{\rm I}\kern-.2em\hbox{\rm P}}}}
\def\IH{{\hbox{{\rm I}\kern-.4em\hbox{\rm H}}}}
\def\ID{{\hbox{{\rm I}\kern-.2em\hbox{\rm D}}}}


\def\Tr{{\rm Tr}\,}
\def\det{{\rm det}}
\def\sv{{\cal v}}


\def\vare{\varepsilon}
\def\barz{\bar{z}}
\def\barw{\bar{w}}

\begin{titlepage}
\begin{flushright}
IP/BBSR/2004-22 \\ 
hep-th/0408184\\
\end{flushright}

\vspace{25mm}
\begin{center}
{\Large {\bf $D=2,\,{\cal N} =2$ Supersymmetric $\sigma$-models on \\   
\vskip 0.4cm
Non(anti)commutative Superspace}  }

\vspace{15mm}
B.~ Chandrasekhar\footnote{chandra@iopb.res.in} 
\\[3mm]
{ Institute of Physics, Bhubaneswar 751 005, INDIA} \\
\end{center}

\thispagestyle{empty}

\vskip2cm


\centerline{\bf ABSTRACT}
\vskip 4mm
\noindent
We extend the results of hep-th/0310137 to show that a
general classical action for $D=2,\,{\cal N}=2$ sigma models 
on a non(anti)commutative superspace is not standard and
contains infinite number of terms,
which depend on the determinant of the non(anti)commutativity 
parameter, $C^{\alpha\beta}$. We show that using 
K\"{a}hler normal coordinates the action can be written in a
manifestly covariant manner.
We introduce vector multiplets and obtain the 
${\cal N}=1/2$ supersymmetry transformations of the theory
in the Wess-Zumino gauge. By explicitly deriving the expressions for
vector and twisted superfields on non(anti)commutative superspace, 
we study the classical aspects of Gauged linear sigma models.
\\
\vspace{2cm}

\today

\end{titlepage}

\newpage

\makeatletter
\rnc{\theequation}{\thesection.\arabic{equation}}
\@addtoreset{equation}{section}
\makeatother
\renewcommand{\thefootnote}{\arabic{footnote}}
\setcounter{footnote}{0}

\tableofcontents
\section{Introduction}                             %

Supersymmetric field theories defined on deformed superspaces have been 
studied for quite some 
time~\cite{Casalbuoni:1975bj}-\cite{Abbaspur:2002xj}. The 
recent interest in such theories is due to the realization that, they
arise naturally in certain limits of string theory, in exactly the 
same way as noncommutative field theories arise in the Seiberg-Witten
low energy limit~\cite{Seiberg:1999vs}. 

In the context of Dijkgraaf-Vafa correspondence~\cite{Dijkgraaf:2002dh},
it was shown that 
the deformation of the algebra of superspace coordinates, allows the 
computation of non-perturbative contributions to the ${\cal N}=1$
superpotential, by summing over certain non-planar diagrams on the
matrix model side.  

To be precise, using the pure spinor approach,
the $D=4$ sigma model action for $D$-branes 
of type II superstring theory
compactified on Calabi-Yau 3-folds was considered
in~\cite{Ooguri:2003qp}.
It was shown that, 
turning on a constant graviphoton background field in four dimensions  
(or more generally 
RR two forms in ten dimensions itself~\cite{deBoer:2003dn}), 
leads to a deformation of correlation functions
of the superspace coordinates as:
\be \label{deformation}
\{ \theta^{\alpha}, \theta^{\beta} \} = 2{\alpha'}^2 F^{\alpha\beta}.
\ee 
Here $\alpha'$ is related to inverse of string tension and $F^{\alpha\beta}$ 
is the self-dual graviphoton field strength. 

Note that the anti-commutation relations of the
remaining superspace coordinates, $\bar{\theta}^{\dot \alpha}$,
are not modified.  This is however,
only possible in a Euclidean space, where setting
the anti self-dual part,  $F^{{\dot\alpha}{\dot\beta}}$ to zero, 
does not affect the string equations of motion. 
Further, it can be 
shown that this configuration
of fluxes is stable and does not back react on the metric, due to
the vanishing of the energy-momentum tensor. 

It was noted that the deformation in 
eqn. (\ref{deformation}) does not survive the field theory limit 
$\alpha' \rightarrow 0$, as long as $F^{\alpha\beta}$ is a constant.
But, the boundary term generated by the graviphoton vertex operator,
survives the field theory limit (as it turns out to be independent of
$\alpha'$) and also breaks half of the supersymmetries.
Nevertheless, it was shown in~\cite{Ooguri:2003qp} that a suitable  
deformation of the gluino anti-commutation relations, cancels the 
boundary term and this restores the ${\cal N}=1$ supersymmetry on
the brane.   
This also restores the standard anti-commutation relations of the
fermionic coordinates $\theta^{\alpha}$ and $\theta^{\beta}$.
Crucial to their analysis were the 
covariant quantization techniques developed 
in~\cite{Berkovits:1994wr}-\cite{Berkovits:1995cb}, for 
studying superstrings in Ramond-Ramond backgrounds in a manifestly
superpoincar\'{e} invariant manner.  

Mechanisms of supersymmetry breaking which come 
from superspace deformation and also
survive in the field theory
limit are very interesting. Hence, for the theory on the brane, the limit 
$\alpha'\rightarrow 0,\:
F^{\alpha\beta}\rightarrow \infty,\: {\alpha'}^2 F^{\alpha\beta} 
= C^{\alpha\beta}=$ fixed, was considered 
in~\cite{Seiberg:2003yz,Berkovits:2003kj}, 
so as to preserve the non-trivial
anti-commutation relations (\ref{deformation}). 
As mentioned before, an important   
consequence of the deformation in eqn. (\ref{deformation}), is that half of the
supersymmetry generators, due to their
dependence on $\theta^{\alpha}$'s become non-linear. As a result, they are 
no more the symmetries of the background. The surviving super translational
symmetry along the ${\bar\theta}$ directions, has been termed as the 
${\cal N}=1/2$ supersymmetry. 

It is useful to note that one still 
continues to use the full superspace, but with the understanding that
the translational symmetry 
in the $\theta$ directions is broken.  Regardless of this
aspect, it was shown in~\cite{Seiberg:2003yz}, that the classical action
of four dimensional ${\cal N}=1$
supersymmetric field theories with the superspace deformation as in 
eqn. (\ref{deformation}), is still Lorentz invariant (in the sense that
the non(anti)commutativity parameter $C^{\alpha\beta}$ appears only
as $(\det\,C)$ in the action). Further,
the F terms were also shown to be invariant
under the surviving ${\cal N}=1/2$ supersymmetry transformations. 

This was followed by number of works on the 
classical and quantum aspects of the Wess
Zumino models in four dimensions, both perturbative and 
non-perturbative. Other interesting 
features~\cite{REY1}-\cite{Grassi:2004ih}
and generalizations to models with ${\cal N}=2$ supersymmetry in
four~\cite{FER}-\cite{Safarzadeh:2004ia}, as well as in two 
dimensions~\cite{SAK,Chandrasekhar:2003uq} have also been discussed. 

In~\cite{Chandrasekhar:2003uq}, we studied the consequences 
of the superspace
deformation (\ref{deformation}) for $D=2,\:{\cal N}=2 $ 
supersymmetric theories characterized by a general K\"{a}hler 
potential and arbitrary superpotential.
The classical 
action was shown to have a power series expansion in the determinant of 
the non(anti)commutativity parameter. The analysis was only
for the case of a single chiral multiplet. Generalization to
include several chiral multiplets and hence to a sigma model, 
is of great importance.

Formulating sigma models in two dimensions is also 
interesting from the target space point of view. 
To be precise, the fact that 
the world-sheet is deformed by the relations (\ref{deformation}),
does not necessarily imply that the target
space shares the same properties. Thus, it is of great 
interest to study ${\cal N}=2$ theories on non(anti)commutative
superspace, with K\"{a}hler manifolds as target spaces.

Further, it is well known that 
${\cal N}=2$ supersymmetric nonlinear sigma 
models~\cite{Zumino:1979et}-\cite{Alvarez-Gaume:hn} 
have provided invaluable 
insights in the physics of lower dimensional systems, 
dynamics of string theory in general curved backgrounds, mirror 
symmetry and Calabi-Yau geometries, topological field theories etc..
These theories are characterized by an underlying K\"{a}hler geometry
which constrains the form of the classical action and, at the 
quantum level, puts severe restrictions on their
ultraviolet structure~\cite{Alvarez-Gaume:1980dk,Alvarez-Gaume:hn}.
Motivated by the above facts, in this paper, we 
first generalize the analysis of~\cite{Chandrasekhar:2003uq}
to study ${\cal N}=2$ supersymmetric non-linear
$\sigma$-models characterized by a K\"{a}hler
potential ${\mathcal K}(\Phi^i,{\bar \Phi}^j)$, where there
are several chiral multiplets, $\Phi^i$, $i = 1,\cdots ,k$. 
This set up 
naturally leads to the formulation of
sigma models with K\"{a}hler manifolds as target spaces. 

On another front,
chiral multiplets in the presence of gauge fields in two dimensions,
have been considered before, for providing
interesting insights into various aspects of Mirror symmetry.
Thus, we extend the analysis to include several
chiral multiplets charged under a single vector multiplet and
study Gauged linear sigma models
(GLSM) on non(anti)commutative superspace. 

As discussed above,
the motivation for studying GLSM's are many fold. First, 
a distinct feature that appears in two dimensions,
compared to four dimensions is that, in addition to the chiral multiplets,
it is also possible to have twisted  
multiplets~\cite{Gates:nk}. Sigma models
having both kinds of multiplets are quite fascinating, as mirror symmetry
interchanges the two. Thus, they 
allow a concrete understanding of the Landau-Ginzburg and Calabi-Yau
phases of ${\cal N}=2$ theories~\cite{Witten:1993yc,Hori:2000kt}.  
Further, mirror symmetry in the presence of fluxes is also being
pursued. The fluxes coming from string theory can either be of
of NS-NS type or RR type. Since, the superspace deformation
in eqn. (\ref{deformation}) is coming from the study of superstrings
in the RR backgrounds, it might be interesting to understand
mirror symmetry in this set up.

Second, Superstring compactifications on Calabi-Yau manifolds 
can generate non-trivial superpotentials in the effective four 
dimensional theory. It is of 
interest to get a better understanding of this superpotential, as it 
encodes important information about the vacuum structure of the theory.
It has been known for a while, that this superpotential can be studied by
looking at the associated sigma model. But, for these sigma models
to be useful, they have to be either conformally invariant or flow
to conformally invariant theories in the IR limit. Nevertheless, it is
still possible to deduce useful information from these models
by twisting them to get topological
theories. The observables and correlation functions in  these
topological sigma models  do not depend on the metric and are 
also invariant under scale transformations.

Due to such varied applications,  
${\cal N} =2$ GLSM's have been studied by many authors. Further, one
can add worldsheets to the boundary by 
putting appropriate boundary conditions on the fields in
the vector multiplet and study $D$-branes 
via these models~\cite{Govindarajan:2000ef}. 
With this motivation, in this paper, we
study the classical aspects of  $D= 2,\:{\cal N} =2$ sigma models
defined on a non(anti)commutative superspace. 

The rest of the paper is organized as follows. In section-2,
we begin with the dimensional
reduction of the relevant formulae from $D=4$ to $D=2$. In the
following subsection, we discuss the various 
supersymmetry multiplets in the 
theory and also explicitly derive the supersymmetry and gauge
transformations in the Wess-Zumino gauge.
In section-3, we discuss the construction of classical action
of the theory, while pointing out the emergence of a series
expansion in $(\det\,C)$. Here, we use a certain normal
coordinate expansion to write the action in a manifestly 
covariant fashion.

In section-4, we present
the classical action of the gauged linear
sigma models, show the invariance under ${\cal N}=1/2$ supersymmetry
transformations and also
make some remarks about the superpotential of the theory. 
We present our conclusions and discussion in section-5.

\section{ ${\cal N} =2$ Superspace and Supermultiplets}                           

In this section, we start by establishing our notations and conventions,
while also reviewing certain general properties of non(anti)commutative
superspace in two dimensions. Some relevant details can also be found
in~\cite{Chandrasekhar:2003uq}, but most of the results have been
rederived so that the connection with four dimensional 
case~\cite{Seiberg:2003yz} is more clear and also to ensure that
the notations are compatible with the ones in~\cite{Witten:1993yc}.
In subsection-2, we discuss the definitions of
the matter and gauge multiplets,  gauge transformation properties
of the component fields and explicitly construct the supersymmetry
transformations in the Wess-Zumino gauge.

Before proceeding, it is useful to mention that we work in 
a Euclidean space, but continue to use Lorentzian signature for 
convenience~\cite{Seiberg:2003yz}. The reason why the underlying
space is Euclidean, can be understood
by going back to the four dimensional relations in 
eqn. (\ref{deformation}).
As discussed before, 
the deformation is imposed 
only over half of the fermionic coordinates, 
while the remaining half still satisfy the same old Grassmannian
algebra. This is only possible in Euclidean space where the 
self-dual component $F_{\alpha\beta}$ can be turned on,
while setting its anti-self dual part $F_{ {\dot \alpha}{\dot \beta} }$
to zero. 

In a Minkowski space-time, the self-dual and (anti)self-dual components 
of the graviphoton field strength are related by
a complex conjugation. However, in a Euclidean space the two 
components transform independently under the two different $SU(2)$
subgroups, which come from 
$SO(4) = SU(2)_L \times SU(2)_R $~\cite{Lukierski:1986jw,TERA}. 

Thus, compared to ${\cal N} = 1$ supersymmetric theories in Minkowski
space-time,  the number of bosonic and fermionic fields of the theory
are doubled and complexified, in the corresponding Euclidean space.
Now, in order to preserve reality conditions, one is
forced to introduce the second supersymmetry. In other words, 
the only way to put
consistent reality conditions on the fields of the theory 
is to extend the superspace to
$D=4, {\cal N} = 2 $ theories. However, we continue to work 
with ${\cal N} = 1$ Euclidean superspace, given the understanding
that all the fields of the theory are complex with 
no reality conditions on them.

\subsection{$D= 4 \rightarrow D =2$ dimensional reduction}         

We now start by discussing the dimensional reduction from $D=4$
to $D= 2$. The superspace coordinates in  $D= 4$ as
given in~\cite{Seiberg:2003yz} are:
$\theta^{\alpha}, {\bar \theta}^{\dot{\alpha}}$ and $y^{\mu}$, where 
$\alpha,\,{\dot \alpha}$ represent the two chiralities of 
spinor indices. Raising and lowering of spinor indices is done
as, $\psi_{\alpha}= \epsilon_{\alpha\beta}\,\psi^{\beta}$, 
$\psi^{\alpha}= \epsilon^{\alpha\beta}\,\psi_{\beta}$, where 
$\epsilon$ is the antisymmetric tensor whose non-zero components 
are given as $\epsilon^{01}= - \epsilon_{10}= 1$. $y^{\mu}$ 
denotes the chiral coordinates and is related to the standard 
$\IR^4$ coordinates as:
\be
y^{\mu} = x^{\mu} + i \theta^{\alpha}\,
\sigma^{\mu}_{\alpha\dot{\alpha}}\,{\bar \theta}^{\dot{\alpha}}.
\ee
The need for using chiral coordinates
can be understood as follows. Once we introduce the deformation:
\be \label{cdef4d}
\{\theta^{\alpha}, \theta^{\beta}\} = C^{\alpha\beta},
\ee 
the standard
$\IR^4$ coordinates $x^{\mu}$, do not commute~\cite{Seiberg:2003yz}. 
However, the coordinates
$y^{\mu}$ can be taken to commute. In fact, all (anti)commutators of
$y^{\mu},\theta^{\alpha}$ and ${\bar \theta}^{\dot{\alpha}} $ vanish,
except (\ref{cdef4d}). 

It will be useful to obtain the $D= 2,\:{\cal N} =2$ superspace by
dimensional reduction of the above formulae, so that later on, the
results obtained here can be directly compared to the ones in $D=4$.

In making the reduction, we take the 2d fields to be independent of $x^1$ 
and $x^2$ and label the fermionic coordinates
as $(\theta^0,\theta^1) = (\theta^-,\theta^+)$ and 
$(\theta_0,\theta_1) = (\theta_-,\theta_+)$. Here, the upper and lower
components are further related as 
$\theta^- = \theta_+ \,,\, \theta^+ = - \theta_-$. Similar
identifications hold for the dotted indices as well. For the 
tensors $\sigma^{\mu}_{\alpha\dot{\alpha}}$ we use~\cite{Wess}:
\bea
\sigma^0 = \pmatrix{ -1 & 0 \cr
              0 & -1 } ,\quad
\sigma^1 = \pmatrix{ 0 & 1 \cr
              1 & 0 }, \quad
\sigma^2 = \pmatrix{ 0 & -i \cr
              i & 0 } ,\quad
\sigma^3 = \pmatrix{ 1 & 0 \cr
              0 & -1 }.
\eea
After dimensional reduction, we find it convenient to
use the following combination of chiral coordinates:
\be \label{chiralbasis}
\xi^- = \frac{1}{2}\,(x^0 - x^3) - i \theta^- {\bar \theta}^- ,
\qquad
\zeta^- = \frac{1}{2}\,(x^0 + x^3) - i \theta^+ {\bar \theta}^+,
\ee
where $\xi^- = \frac{1}{2}\,(y^0 - y^3) $ and 
$\zeta^- = \frac{1}{2}\,(y^0 + y^3)$. Our non(anti)commutative
superspace can be obtained by translating the relations (\ref{cdef4d}),
to $D=2$ as:
\be \label{Cdeformation}
\{\theta^-,\theta^- \}= C^{00}, \qquad   
\{\theta^-,\theta^+ \} = C^{01},\qquad
\{\theta^+,\theta^- \} = C^{10},\qquad 
\{\theta^+,\theta^+ \} =   C^{11}.
\ee
Functions of $\theta^-$ and $\theta^+$, say $f(\theta^-,\theta^+)$
and  $g(\theta^-,\theta^+)$
are Weyl-ordered using the following definition 
of star product:
\bea \label{star1}
f*g &=& f\:
\exp\left(
-{C^{00}\over 2 }\overleftarrow{\partial_{\theta^-} }
\overrightarrow{\partial_{\theta^-} }
-{C^{01}\over 2 }\overleftarrow{\partial_{\theta^-} }
\overrightarrow{\partial_{\theta^+} } 
- {C^{10}\over 2 }\overleftarrow{\partial_{\theta^+} }
\overrightarrow{\partial_{\theta^-} }
-{C^{11}\over 2 }\overleftarrow{\partial_{\theta^+} }
\overrightarrow{\partial_{\theta^+} }\right)\:g.
\eea

The generators of supersymmetry transformations, written in 
the chiral basis (\ref{chiralbasis}) are,
\be \label{Q}
Q_{\pm} = -\frac{\partial}{\partial\theta^{\pm}}, \qquad 
\bar{Q}_{\pm} = -\frac{\partial}{\partial{\bar\theta}^{\pm}} 
- 2 i \theta^{\pm}
\left(\frac{\partial}{\partial y^0}
\pm \frac{\partial}{\partial y^3} \right),
\ee
and these anti-commute with the remaining set of derivatives,
written in chiral basis as:
\be \label{derivatives}
D_{\pm} = \frac{\partial}{\partial\theta^{\pm}} 
- 2 i {\bar\theta}^{\pm}
\left(\frac{\partial}{\partial y^0}
\pm \frac{\partial}{\partial y^3} \right), \qquad
\bar{D}_{\pm} = -\frac{\partial}{\partial{\bar\theta}^{\pm}}\, .
\ee
In the chiral basis (\ref{chiralbasis}), 
the algebra of the supercovariant derivatives (\ref{derivatives}),
does not get modified due to the deformation (\ref{Cdeformation}), as
seen below:
\be \label{dalgebra}
\{{\bar D}_{\pm},D_{\pm} \}= 2i\left(\frac{\partial}{\partial y^0}
\pm \frac{\partial}{\partial y^3}\right) \qquad {\rm ~and~rest~all~zero~}.
\ee
However, the algebra of supercharges given in eqn. (\ref{Q}) gets
modified:
\bea \label{qalgebra}
\{Q_{\pm}, {\bar Q}_{\pm} \} &=& -2i \left(\frac{\partial}{\partial y^0}
\pm \frac{\partial}{\partial y^3}\right) \xx
\{{\bar Q}_-, {\bar Q}_- \} &=& -4\,C^{00}\left(\frac{\partial}{\partial y^0}
- \frac{\partial}{\partial y^3}\right)^2 \xx
\{{\bar Q}_+, {\bar Q}_+ \} &=& -4\,C^{11}\left(\frac{\partial}{\partial y^0}
+ \frac{\partial}{\partial y^3}\right)^2 \xx
\{{\bar Q}_-, {\bar Q}_+ \} &=& 
- 4\,C^{01}\left(\frac{\partial^2}{({\partial y^0})^2}
- \frac{\partial^2}{({\partial y^3})^2}\right),
\eea
and rest all zero. As stated before, due to the dependence of
$\bar Q$'s on the non(anti)commutative coordinates $\theta^{\pm}$,
it is no more a symmetry of the theory. From the algebra (\ref{qalgebra}),
the only unbroken symmetry generators are $Q_{\pm}$. Hence,
we only use these ${\cal N}=1/2$ supersymmetry generators to 
study the theory.

\subsection{${\cal N} =2$ multiplets} 

Let us start by discussing the ${\cal N} =2$ matter and gauge multiplets
in two dimensions. For the $C=0$ case, the results are 
summarized in~\cite{Witten:1993yc}. For the case with $C\neq 0$, the
definition of the vector superfield and the subtleties in defining
gauge transformations in $D=4$ has been discussed in~\cite{Seiberg:2003yz}.
The discussion has been further extended to include chiral multiplets
in~\cite{ARAKI}. Thus, the simplest way to obtain the vector and chiral 
multiplets in two dimensions is to do a dimensional reduction of the 
relevant formulae given in four dimensions. 

As we will see, a naive dimensional reduction may not show some 
critical aspects associated with the definition of the multiplets. 
Thus, we choose to derive the proper definitions of vector and 
chiral superfields in $D=2$ for the case $C\neq 0$.  Later on,
we compare these definitions with the ones obtained by a dimensional
reduction and point out the differences. What we will see is that,
a direct reduction of the definition of Vector superfields 
from $D=4$ may
give some additional terms, which can be ignored in $D=2$.

\begin{flushleft}
{\underline { Vector Multiplet}}
\end{flushleft}
\vskip 0.2cm

Since, one of our interests is
in formulating a gauge theory, we first introduce the Vector superfield
$V$. For simplicity, in this work we only consider abelian gauge groups,
in which case $V$ is a single real function on the superspace. Towards
the end, we comment on the generalization to the case of non-abelian
gauge groups. 

Even
after imposing the reality condition, there
is a residual gauge invariance under which the vector superfield 
transforms infinitesimally as:
\be \label{gaugetr}
\delta e^{V} = -i{\bar \Lambda}*e^{V} + i e^{V}*\Lambda\, ,
\ee
where $\Lambda = -\alpha(\xi^-,\zeta^-)$ and 
${\bar \Lambda}=  -\alpha(\xi^+,\zeta^+)$ are the gauge 
parameters with $\xi^+ = \xi^- + 2i \theta^- {\bar \theta}^-$ and
$\zeta^+ = \zeta^- + 2i \theta^+ {\bar \theta}^+$.
This residual 
gauge invariance can be partially fixed by going to a Wess-Zumino gauge,
in which case, $V$ takes the form:
\bea \label{vector}
V_{\rm wz}
&=& 
-{\bar\theta}^-\,\theta^- \,\nu_{\xi}
-{\bar\theta}^+\,\theta^+\, \nu_{\zeta}\,
+\,\sqrt{2}\,{\bar\theta}^+\,\theta^-\,\sigma \,
+\,\sqrt{2}\,{\bar\theta}^-\,\theta^+\, \bar {\sigma}\,
+\, 2i\,\theta^-\,\theta^+\,(\,{\bar\theta}^+\,{\bar \lambda}_+ \, 
+\,{\bar\theta}^-\,{\bar \lambda}_-\,) \,\xx 
&-&\, 2i\,{\bar\theta}^-\,{\bar\theta}^+\,(\, \theta^-\,\lambda_-\,
+\,\theta^+ \,\lambda_+\,) \, 
-\,2\,\theta^- \theta^+{\bar \theta^-}{\bar \theta}^+\,
(\,D \,+\,\frac{i}{2}\partial_{\zeta^-}\nu_{\xi} 
+\,\frac{i}{2}\partial_{\xi^-}\nu_{\zeta} \,).
\eea
In the above definition of the vector multiplet, 
for gauge fields $\nu_0,\nu_1$, we have introduced the notation
$\nu_{\xi}= (\nu_0 - \nu_1)\,$ and 
$\nu_{\zeta} = (\nu_0 + \nu_1)$, as this combination will occur quite 
frequently in chiral basis. Further, in eqn. (\ref{vector}) 
$\sigma,\,{\bar \sigma}$ are complex
scalars, $\lambda_{\pm},{\bar\lambda}_{\pm}$ are the gauginos and
$D$ is an auxiliary field.

To find out the gauge transformation properties of the component
fields, we write ${\bar \Lambda}$ in terms of $(\xi^-,\zeta^-)$ 
coordinates as:
\be
{\bar \Lambda} = - \alpha - 2i \theta^- \bar \theta^-
\partial_{\xi^-} \alpha - 2i \theta^+ \bar \theta^+ 
\partial_{\zeta^-} \alpha
- 4\theta^- {\bar \theta}^-\theta^+ \bar \theta^+ 
\partial_{\xi^-}\partial_{\zeta^-} \alpha,
\ee
and calculate R.H.S. of eqn. (\ref{gaugetr}), where for V, we use
the definition derived in eqn. (\ref{vector}). 
Some
terms in the calculation, namely the ones depending on $C$, are given 
below (identities used in 
the calculation are given in Appendix) :
\bea \label{rhs}
-i{\bar \Lambda}*e^{V} + i e^{V}*\Lambda 
&=& \:\: \,\bar \theta^- \bar \theta^+  \left[
i\, \partial_{\xi^-} \alpha\,
\left(\,\theta^+ C^{00} + \theta^- C^{01}\,\right){\bar \lambda}_+ 
\,+\,
\partial_{\zeta^-} \alpha
\,\left(\,\theta^- C^{11} + \theta^+ C^{10}\,\right)
{\bar \lambda}_- \right. \xx
&+& \left. \,2\,
( -\sqrt 2 C^{00} \sigma \partial_{\xi^-} \alpha 
+ C^{01} \nu_{\zeta}\partial_{\xi^-} \alpha 
-  C^{10}\nu_{\xi} \partial_{\zeta^-} \alpha 
+ \sqrt 2  C^{11} {\bar \sigma}\partial_{\zeta^-} \alpha ) \right].
\eea
Now, comparing the variation of the vector superfield and the
result in eqn. (\ref{rhs}), one can directly obtain
the gauge transformations of the component fields of the
vector multiplet, as given below:
\bea
\delta_g \nu_{\xi} &=& - 2\partial_{\xi^-}\alpha \xx
\delta_g \nu_{\zeta} &=& - 2\partial_{\zeta^-}\alpha \xx
\delta_g (\sigma, \bar \sigma) &=& 0 \xx 
\delta_g D &=& 0 \xx
\delta_g {\bar \lambda}_{\pm} &=& 0 \xx
\delta_g \lambda_- &=& 
- \frac{i}{2}\left(\,C^{01}{\bar \lambda}_+  \partial_{\xi^-}\alpha
+  \,C^{11} {\bar \lambda}_-  \partial_{\zeta^-}\alpha \right)  \xx
\delta_g \lambda_+  &=& 
- \frac{i}{2}\left(\,C^{00}{\bar \lambda}_+  \partial_{\xi^-}\alpha
+  \,C^{10} {\bar \lambda}_-  \partial_{\zeta^-}\alpha \right).
\eea
These are not the standard gauge transformation properties of the 
component fields, due
to the new $C$-dependent terms present in 
$\delta_g \lambda_{\pm}$. However,
as suggested in~\cite{Seiberg:2003yz}, it is possible to 
cancel the new terms seen in $\delta_g \lambda_{\pm}$ , by
modifying the definition of $V_{\rm wz}$ to include certain  
new $C$-dependent terms. In fact from eqn. (\ref{gaugetr}), 
it is possible to guess the kind
of terms that need to be added to $V_{\rm wz}$. 
The new terms to be added are of the following kind:
\be  \label{Vc}
V_c =  i{\bar\theta}^-{\bar\theta}^+\left[
\theta^- ( C^{01} {\bar \lambda}_+ \nu_{\xi} +  C^{11}{\bar \lambda}_-
\nu_{\zeta} ) + \theta^+ (  C^{00} {\bar \lambda}_+ \nu_{\xi} + 
 C^{10}{\bar \lambda}_- \nu_{\zeta} )\right].
\ee
Below we argue, that modifying the definition of vector 
superfield as in eqn. (\ref{vmodified}), has the effect of 
canceling the first two terms in the quantity given in 
eqn. (\ref{rhs}). This in turn corresponds to restoring 
the standard gauge transformation
property of the gauginos, i.e., $\delta_g \lambda_{\pm}=0$. 

The way to guess the new terms given in eqn. (\ref{Vc}), is
to note that $\partial_{\xi^-}\alpha$ and 
$\partial_{\zeta^-}\alpha $
appearing in $\delta_g \lambda_{\pm}$ are nothing but the
gauge transformations of the gauge fields $\nu_{\xi}$ 
and $\nu_{\zeta}$. Thus,  the terms in $V_c$ have been 
chosen in such a way that,  $\delta_g V_c$ looks similar
to the terms appearing in $\delta_g \lambda_{\pm} $. 
The rest is to adjust the coefficients by making this ansatz..

The remaining terms in the second line of eqn. (\ref{rhs}), can
also be understood to be coming from a modification of
the gauge parameter as shown below:
\bea \label{gparameter}
{\bar \Lambda} &=& - \alpha - 2i \theta^- \bar \theta^-
\partial_{\xi^-} \alpha - 2i \theta^+ \bar \theta^+ 
\partial_{\zeta^-} \alpha
- 4\theta^- {\bar \theta}^-\theta^+ \bar \theta^+ 
\partial_{\xi^-}\partial_{\zeta^-} \alpha \xx
&+& - i
\,{\bar\theta}^-\,{\bar\theta}^+\,
\left[ \,
-\sqrt 2 C^{00} \sigma \partial_{\xi^-} \alpha 
+ C^{01} \nu_{\zeta}\partial_{\xi^-} \alpha 
-  C^{10}\nu_{\xi} \partial_{\zeta^-} \alpha 
+ \sqrt 2  C^{11} {\bar \sigma}\partial_{\zeta^-} \alpha
\right].
\eea
To summarize, choosing the 
final form of vector superfield in the $C$-deformed case to be,
\be \label{vmodified}
V^c_{\rm wz} = V_{\rm wz} + V_c,
\ee
with $V_c$ given as in eqn. (\ref{Vc}) and modifying the
gauge parameter as in eqn. (\ref{gparameter}), the standard gauge
transformation properties of the component fields are restored.

We note that the additional terms added to the definition of the
vector superfield in four dimensions~\cite{Seiberg:2003yz} are
a bit different from the ones given in eqn. (\ref{Vc}). If we
dimensionally reduce the definitions given 
in~\cite{Seiberg:2003yz}, we get terms of the kind:
\be 
i{\bar\theta}^-{\bar\theta}^+
\theta^- (C^{01} {\bar \lambda}_+ \sigma +  C^{11}{\bar \lambda}_-
{\bar \sigma}), 
\ee
which may contribute to  eqn. (\ref{Vc}). 
However, these terms contain 2d scalars (coming from
4d gauge fields) which do not vary under gauge transformations 
in $D=2$ and hence, do not affect
the gauge transformation properties of any of the component fields.
Thus, these terms do not play any role in the present analysis.
Further, in~\cite{Seiberg:2003yz}, the vector superfield was
apriori assumed to be matrix valued and the theory was  
non-abelian. Since, for the present case, we only consider
abelian gauge groups, these terms do not occur. However, it 
is useful to note that, if there are several vector multiplets,
then there is a restriction on the allowed gauge groups in the
theory~\cite{TERA}.

Before proceeding, it will be useful to write down the
powers of the vector superfield (\ref{vmodified}), as shown below:
\bea \label{vpowers}
V_*^2 &=& V * V \xx
&=& 2 {\bar\theta}^-{\bar\theta}^+\left [ \theta^-\theta^+
(-\nu_{\zeta}\nu_{\xi}+ 2 \sigma {\bar \sigma} ) -
(\det C) {\bar \lambda}_- {\bar \lambda}_+ \, \right], \xx
V_*^3 &=& 0.
\eea
One can see that, as in the standard $C=0$ case,
star product of more than two Vector superfields 
vanishes~\cite{Seiberg:2003yz}, and this will be needed while
writing down the action.

\begin{flushleft}
{\underline { Twisted Multiplets}}
\end{flushleft}

It has been known for quite some time that ${\cal N}=2$  
sigma models having both chiral and twisted chiral 
multiplets are helpful in understanding Mirror
symmetry. Hence, for the present case, 
we follow~\cite{Witten:1993yc} and construct the
twisted chiral superfield for an abelian gauge theory 
as~\cite{Rocek:1991ps,Witten:1993yc}:
\be
\Sigma = \frac{1}{\sqrt 2}\, {\bar D}_+\,D_-\: V\, ,
\ee
where the modified Vector superfield V is defined 
in eqn. (\ref{vmodified}). Using the algebra of 
the supercovariant derivatives given in  
eqn. (\ref{dalgebra}), it is possible to show 
that the twisted chiral superfield satisfies the conditions,
$D_- \Sigma = 0,\:{\bar D}_+ \Sigma = 0$ and can be 
written in terms of its components as:
\bea \label{twistedC}
\Sigma &=& \sigma +  \,i\sqrt 2  \theta^+ {\bar \lambda}_+ 
+\,i\sqrt 2 \,{\bar\theta}^-
\left[\, - \lambda_- + \frac{1}{2}   
C^{01} \nu_{\xi} \,(\,{\bar \lambda}_+ 
+ 2i\theta^+{\bar\theta}^+ \partial_{\zeta^-}{\bar \lambda}_+ \,)
+ \frac{1}{2} C^{11}\nu_{\zeta} \,( \,{\bar\lambda}_-  
+ 2i\theta^+{\bar\theta}^+ \right. \xx
&\times& \left. \partial_{\zeta^-}{\bar \lambda}_-
\,) \,\right] 
- \sqrt 2 \,{\bar\theta}^- \theta^+ 
(  D - \frac{i}{2} \nu_{\xi\zeta} ) 
-\, 2i {\bar\theta}^-\theta^- \partial_{\xi^-} \sigma 
- 2\sqrt 2 {\bar\theta}^- ( \theta^+ \theta^- - \frac{1}{2}C^{10})\,
 \partial_{\xi^-}{\bar \lambda}_+ \,,
\eea
where $\nu_{\xi\zeta}= \partial_{\xi^-}\nu_{\zeta} 
- \partial_{\zeta^-}\nu_{\xi}  $ is the gauge field strength.
Twisted anti-chiral superfield satisfying 
$D_+ {\bar \Sigma} = 0$ and ${\bar D}_- {\bar \Sigma} = 0$
can be obtained in an analogous way from,
${\bar\Sigma} = \frac{1}{\sqrt 2}\, {\bar D}_-\,D_+\: V$, and is
given below:
\bea \label{twistedAC}
{\bar \Sigma} &=& {\bar \sigma} -\, i \sqrt 2  \theta^- {\bar \lambda}_- 
+\,i\sqrt 2 \,{\bar\theta}^+
\left[\,  \lambda_+ - \frac{1}{2}  
C^{00}\nu_{\xi} \,(\, {\bar \lambda}_+ 
+ 2i\theta^-{\bar\theta}^- \partial_{\xi^-}{\bar \lambda}_+ \,)
- \frac{1}{2} C^{10}\nu_{\zeta}\, (\, {\bar\lambda}_-  
+ 2i\theta^-{\bar\theta}^- \right. \xx 
&\times& \left.\partial_{\xi^-}{\bar \lambda}_- \, ) \,\right] 
- \sqrt 2 \,{\bar\theta}^+ \theta^- 
(  D + \frac{i}{2} \nu_{\xi\zeta}  ) 
-\, 2i {\bar\theta}^+\theta^+ \partial_{\zeta^-} {\bar \sigma} 
+ 2\sqrt 2 {\bar\theta}^+(\theta^- \theta^+ - \frac{1}{2}C^{01})\,
 \partial_{\zeta^-}{\bar \lambda}_- \,.
\eea
All the component fields of twisted superfields are functions of
$(\xi^-, \zeta^-)$. It is useful to compare the definitions of
twisted superfields given in eqns. (\ref{twistedC})  and 
(\ref{twistedAC}) with the ones given in~\cite{Witten:1993yc}. 
The only difference is the new $C$-dependent terms, some of which arise
from the additional terms added to the definition of vector 
superfield. These terms have also been expanded around 
$(\xi^-, \zeta^-)$ coordinates.
Other $C$-dependent terms, for instance, the term
in the second line of eqn. (\ref{twistedC}) can be obtained from the
twisted chirality condition. 

A vector superfield by itself is not a gauge 
invariant object and hence, is not
directly used to construct the action for the gauge fields.
Rather, the twisted superfields derived from $V$ 
are used in writing down a gauge invariant 
action for gauge fields. In other words, twisted superfields
play the role of gauge invariant field strength for the superspace
$U(1)$ gauge fields.

We now write down the supersymmetry transformations of the 
component fields of the vector multiplet. It is easier to
derive them from the twisted multiplets as follows:
\be
\delta \Sigma = \left( \epsilon^+ Q_+ 
+ \epsilon^- Q_-  \right)\:\Sigma ,
\ee
with similar relations 
for the twisted anti-chiral multiplet.
Comparing the right hand side of the 
above equation with the variation of the component
fields in the definition of $\Sigma$ given 
in eqn. (\ref{twistedC}), we get:
\bea
\delta \sigma &=& i \sqrt 2 \epsilon^+ {\bar \lambda}_+ \xx
\delta {\bar\sigma} &=& - i \sqrt 2 \epsilon^- {\bar \lambda}_- \xx
\delta {\bar \lambda}_+ &=& 0 \xx
\delta {\bar \lambda}_-  &=& 0 \xx
\delta \nu_{\zeta} &=& 2i \epsilon^- {\bar \lambda}_+ \xx
\delta \nu_{\xi} &=& -2i \epsilon^+ {\bar \lambda}_-\,.  
\eea
The above transformations are same even for the $C=0$ theory. 
However, the transformation properties of the 
remaining component fields get modified by certain new 
terms, as seen below:
\bea \label{lambdasusy}
\delta \lambda_+ &=& - \sqrt 2 \epsilon^+ 
\partial_{\zeta^-}{\bar\sigma} 
+ i \epsilon^- ( D + \frac{i}{2} \nu_{\xi\zeta}) 
+ i ( C^{00} \epsilon^+ + C^{10} \epsilon^-)
 {\bar \lambda}_+{\bar \lambda}_-  \xx
\delta \lambda_- &=& \sqrt 2 \epsilon^+ 
\partial_{\xi^-}\sigma 
- i \epsilon^+ ( D - \frac{i}{2} \nu_{\xi\zeta}) 
+ i ( C^{01} \epsilon^+ + C^{11} \epsilon^-)
 {\bar \lambda}_+{\bar \lambda}_- \, .
\eea
It is useful the compare the above results with the ones obtained
by dimensional reduction from~\cite{Seiberg:2003yz}.

\begin{flushleft}
{\underline {Chiral multiplets}}
\end{flushleft}
Now, the Chiral and anti-Chiral superfields satisfying 
${\bar D}_{\pm} \Phi =0$ and 
$D_{\pm} \bar{\Phi}_o =0$ 
respectively, can be written in a Weyl ordered form,
as shown below~\cite{Chandrasekhar:2003uq,Seiberg:2003yz}:
\bea \label{cc}
\Phi
&=& \phi
+ \sqrt 2 \,\theta^- \psi_- 
+ \sqrt 2 \,\theta^+ \psi_+
- 2 \,\theta^-\theta^+\,F , \\
\label{ac}
{\bar \Phi}_o
&=& \bar{\phi} - \sqrt 2\, {\bar\theta}^- {\bar \psi}_-
- \sqrt 2 \, {\bar\theta}^+ {\bar \psi}_+ + 2 i \theta^-{\bar\theta}^-
\partial_{\xi^-}\bar{\phi} + 2i\theta^+{\bar\theta}^+ 
\partial_{\zeta^-}\bar{\phi} \xx
&+&{\bar\theta}^-{\bar\theta}^+\left( \,2\bar{F}
-\,2\sqrt 2i \theta^-\partial_{\xi^-}{\bar \psi}_+
+\,2\sqrt 2i \theta^+\partial_{\zeta^-}{\bar \psi}_-
+ 4 \, \theta^-\theta^+ \,\partial_{\xi^-}
\partial_{\zeta^-} \bar{\phi} \, \right).
\eea
Note that we have used $C^{01}= C^{10}$ in Weyl ordering the 
above expressions. Also, 
the definitions given in eqn. (\ref{chiralbasis}) have been
used in writing the anti-chiral superfield. All the component
fields are taken to be functions of $\xi^-$ and $\zeta^-$,
unless specified otherwise.

The ${\cal N} =1/2$ supersymmetry transformations 
of the component fields in the chiral and anti-chiral multiplet
are standard and 
were also derived in~\cite{Chandrasekhar:2003uq}. We give them below 
for later use:
\bea \label{susy}
\delta \phi &=& \sqrt 2 \epsilon^+ \psi_+ + 
\sqrt 2 \epsilon^- \psi_- \xx
\delta\psi_- &=& - \sqrt 2 \epsilon^+ F \xx
\delta\psi_+ &=&  \sqrt 2 \epsilon^- F \xx
\delta F  &=& 0 \xx
\label{susybar}
\delta {\bar \phi} &=& 0 \xx
\delta {\bar \psi}_+ &=&  i \sqrt 2 \epsilon^+ 
\partial_{\zeta^-}{\bar \phi} \xx
\delta {\bar \psi}_- &=&  i \sqrt 2 \epsilon^- 
\partial_{\xi^-}{\bar \phi} \xx
\delta {\bar F} &=&  i \sqrt 2 \epsilon^+ 
\partial_{\zeta^-}{\bar \psi}_- -  i \sqrt 2 \epsilon^- 
\partial_{\xi^-}{\bar \psi}_+ \,.
\eea
Note that the above transformations do not take into account 
the coupling with the vector multiplet. 
Supersymmetry transformations for the matter multiplet coupled
to the gauge multiplet will be derived explicitly in the following
subsection.

Now we couple the matter and the vector multiplets by making the 
chiral and anti-chiral superfields transform in a certain
representation of the gauge group. Thus, under a gauge
transformation, the matter superfields transform as 
$~\Phi'\,=\, e^{-i\Lambda}\,*\, \Phi\:,~  {\bar \Phi}'_o\,=\, 
{\bar \Phi}_o\,*\,e^{i{\bar \Lambda}}~$ or infinitesimally as:
\be \label{chiralgauge}
\delta \Phi = - i\Lambda * \Phi, \qquad 
\delta {\bar \Phi_o} = i {\bar \Phi_o} * {\bar \Lambda},
\ee
where $\Lambda = - \alpha$ and 
${\bar \Lambda}$ gets modified due to the additional $C$-dependent
terms added to the vector superfield as given in eqn. (\ref{gparameter}).

Following the discussion in the case of the vector multiplet, one
can compare the L.H.S. and R.H.S. of each of the equations 
in (\ref{chiralgauge}), to
get the gauge transformation properties of the (anti)chiral multiplet.
As it turns out and also pointed out in~\cite{ARAKI}, the component
fields of the matter multiplet do not have standard transformation
properties, due to certain additional $C$-dependent terms. For 
instance, the transformation of the auxiliary field takes the form:
\bea
\delta_g {\bar F} &=& -i \alpha  {\bar F} 
+  \frac{ {\bar \alpha}}{2}\left[ \,C^{10}\nu_{\zeta} 
\partial_{\xi^-} \alpha 
- \sqrt 2 C^{00} \sigma \partial_{\xi^-}\alpha 
+  \sqrt 2 C^{11}{\bar \sigma}
\partial_{\zeta^-}\alpha 
- C^{01} \nu_{\xi} \partial_{\zeta^-}\alpha  \, \right] \xx
&+&i\, C^{01} \partial_{\xi^-}{\bar \phi}
\partial_{\zeta^-}\alpha
- i\, C^{10} \partial_{\zeta^-}{\bar \phi} 
\partial_{\xi^-} \alpha.
\eea
However, as discussed in
the case of the vector multiplet above, it is possible to 
guess the terms that should be added to the chiral superfields,
so that the component fields have the standard gauge transformation
properties. Thus, we  modify the definition of the  
anti-chiral superfield by adding certain $C$-dependent terms 
as $~{\bar \Phi} = {\bar \Phi_o} + {\bar \Phi_c}~$, where
${\bar \Phi_o}$ is defined in eqn. (\ref{ac}) and ${\bar \Phi_c}$ is given as:
\be \label{phiC}
{\bar \Phi_c} = -i\,{\bar\theta}^-{\bar\theta}^+
\left[ \sqrt 2\,C^{00}\, \partial_{\xi^-}(\sigma\,{\bar \phi})
\,-\,\sqrt 2\,C^{11}\, \partial_{\zeta^-}({\bar \sigma}\,{\bar \phi})
\,-\,  C^{10} \partial_{\zeta^-}(\nu_{\xi} {\bar \phi})
\,+\,  C^{01} \partial_{\xi^-}(\nu_{\zeta} {\bar \phi})\,
\right].
\ee
One can again check that, the new terms ${\bar \Phi_c}$ are such 
that, the $C$-dependent terms appearing in $\delta_g {\bar F}$ 
are canceled. It turns out that 
the definition of the chiral superfield need not be modified.
With the modified definitions of the matter superfields, we write 
down the gauge transformations of the component fields as shown below:
\bea \label{gtrmatter}
\delta_g \phi &=& + i \alpha \phi, \qquad
\delta_g \psi_{\pm}  = + i \alpha \psi_{\pm}, \qquad
\delta_g F = + i \alpha F , \\
\delta_g {\bar \phi} &=& -i \alpha {\bar \phi}, \qquad
\delta_g  {\bar \psi}_{\pm} = -i \alpha {\bar \psi}_{\pm}, \qquad
\delta_g {\bar F} = - i \alpha {\bar F} .
\eea
It is important to note that the additional $C$-dependent terms 
that have been added to the anti-chiral superfield do not spoil
the chirality conditions. Thus, the new field ${\bar \Phi}$ still 
satisfies $D_{\pm}{\bar \Phi} =0$.

\subsection{Wess-Zumino Gauge }                                    

Supersymmetry transformations for the component fields of the vector
multiplet, in the Wess-Zumino(WZ) gauge were derived in section-2. Here,
we discuss the supersymmetry transformation properties of the 
chiral multiplet.

It is well known that the
WZ-gauge breaks supersymmetry. In other words, the supersymmetry
transformations do not leave the gauge-fixing conditions invariant. 
For this reason, in the WZ-gauge, every supersymmetry transformation
has to be supplemented by an appropriate gauge transformation.
The supersymmetry transformations of the chiral and anti-chiral
multiplets are already given in eqns. (\ref{susy}). 

Before doing anything, one can guess that
the ${\cal N}=1/2$ supersymmetry transformations of the chiral multiplet
remain unchanged even after the coupling with 
vector multiplet. This can be understood by noting that the 
modification for the chiral multiplet comes from variations 
under ${\bar \epsilon}\,{\bar Q}$. However,
as discussed earlier, ${\bar Q}$'s are no more the symmetries
of the theory and hence, the supersymmetry transformations of
chiral multiplet do not change and are same as the ones given in
eqn. (\ref{susy}).
However, the supersymmetry transformations of the anti-chiral multiplet
get modified in the WZ-gauge and we derive them below. 

There are various ways to realize the supersymmetry transformations
in the WZ-gauge. The straightforward way to the derive the 
transformations is note that, in the presence of gauge fields,
the anti-chiral superfield takes the form:
\be \label{newphi}
{\bar \Phi}'= {\bar \Phi}*e^V . 
\ee
Since, in this work we only consider a single vector multiplet,
the superfield ${\bar \Phi}'$ transforms under a $U(1)$ gauge 
group and satisfies the condition 
${\cal D}_{\pm} {\bar \Phi}' = 0$, where ${\cal D}_{\pm}$ denotes
a gauge covariant derivative (the explicit form of which we 
introduce later). 
The right hand side of the eqn. (\ref{newphi}) can be evaluated 
straightforwardly. 
Then, one can calculate $\delta {\bar \Phi}' = (\epsilon^+ Q_+ 
+ \epsilon^- Q_-)\, {\bar \Phi}'$ and compare it with the variation of
the right hand side of eqn. (\ref{newphi}).  

The above procedure will give the combined 
supersymmetry and gauge transformations of the component fields.
We will however, resort to another method by which one can
calculate the appropriate gauge transformation corresponding
to every supersymmetry transformation. Since, we have already
calculated the supersymmetry transformations of the anti-chiral
multiplet in eqn. (\ref{susy}), all we need to do is to 
determine the appropriate gauge transformations. We follow
the method discussed in~\cite{PREM}.

A general Vector superfield on a non(anti)commutative 
superspace can be written as:
\be \label{V}
V\: = \: V^c_{\rm wz} \:+\: i\,(\,{\bar {\tilde \Lambda}} 
- {\tilde \Lambda} \,),
\ee
where the fields which survive in the WZ-gauge and the other
fields which can be set equal to zero have been separated 
out in eqn. (\ref{V}). Here,
$V^c_{\rm wz}$ is the vector superfield in the WZ-gauge, as 
given in eqn. (\ref{vmodified}) 
and (${\bar {\tilde \Lambda}}$) ${\tilde \Lambda}$ 
is the (anti)chiral superfield containing other fields,
as shown below: 
\bea \label{wzcc}
{\tilde \Lambda}
&=& {\tilde \phi}
+ \sqrt 2 \,\theta^-  {\tilde\psi_- }
+ \sqrt 2 \,\theta^+  {\tilde\psi_+}
- 2 \,\theta^-\theta^+\, {\tilde F} , \\
\label{wzac}
{\bar {\tilde \Lambda}}
&=&  {\tilde {\bar \phi}} - \sqrt 2\, {\bar\theta}^-  {\tilde{\bar \psi}}_-
- \sqrt 2 \, {\bar\theta}^+  {\tilde{\bar \psi}}_+ + 
2 {\bar\theta}^-{\bar\theta}^+\, {\tilde {\bar F}}. 
\eea
For the rest of the analysis, 
we set all the component fields of ${\tilde \Lambda}$
in eqn. (\ref{wzcc}) to zero. This is consistent with the WZ
gauge choice due the reasons already discussed above.
Now, if one naively sets all the component fields appearing in
eqn. (\ref{wzac}) to zero, then that is not
enough to preserve the gauge choice. This is due to the 
fact that some of the component fields may transform under 
${\cal N} = 1/2$ supersymmetry transformations. As a result,
the fields which have been set equal to zero, can be recovered
back by a supersymmetry transformation.

Thus, for the anti-chiral multiplet,
one can make a choice for the component fields of 
${\bar {\tilde \Lambda}}$
appearing in eqn. (\ref{wzac}). For some of the fields,
the choice does not involve any $C$-dependent pieces and
they are already known in the standard literature. For
instance, for some of the fields one can guess the terms
by looking at the analogous expressions given in~\cite{PREM},
for the $C=0$ case in four dimensions. Thus, we choose:
\bea
\bar{{\tilde \phi}} &=& 0, \xx
 {\tilde{\bar \psi}}_- &=& \sqrt 2\epsilon^-\nu_{\xi} {\bar \phi} 
- 2 \epsilon^+ {\bar \sigma} {\bar \phi}  , \xx
 {\tilde{\bar \psi}}_+ &=&\sqrt 2\epsilon^+\nu_{\zeta} {\bar \phi} 
- 2 \epsilon^- \sigma {\bar \phi}. 
\eea
For the auxiliary field $\bar {\tilde F}$, the choice involves
adding certain $C$-dependent pieces apart from the usual 
pieces. There is a way to guess the terms, but what we will do
is to give the relevant terms below and then at the end, it
will be clear as to why this particular choice has been 
made:
\be \label{tildeF}
\bar{{\tilde F}} =  
   2\, C^{01}  \epsilon^-{\bar \lambda}_+ \nu_{\xi} 
+  2 \,C^{11} \epsilon^- {\bar \lambda}_- \nu_{\zeta}  
+  2 \,C^{00}\epsilon^+ {\bar \lambda}_+ \nu_{\xi}  
+  2 \, C^{10}\epsilon^+{\bar \lambda}_- \nu_{\zeta}  .
\ee
Note that $\bar{{\tilde F}}$ will have some $C=0$ pieces as well.
Further,  
the gauge parameter has been chosen in such a way that the sum of
a supersymmetry and a gauge transformation vanishes, i.e., 
$(\delta_s + \delta_g) {\bar {\tilde \Lambda}} = 0$.

Hence, the sum of supersymmetry and gauge transformations
for the component fields of the anti-chiral multiplet, in 
the WZ gauge can now be calculated. 
The ones which remain same, as in the $C=0$ theory
are given below:
\bea \label{susySG}
(\delta_s + \delta_g ) {\bar \phi} &=& 0,   \xx
(\delta_s + \delta_g ) {\bar \psi}_{-} &=&  i\sqrt 2
\,\epsilon^- {\bar {\cal D}}_{\xi^-}{\bar \phi} 
-  2 Q \epsilon^+ \bar{\sigma }{\bar \phi},      \xx
(\delta_s + \delta_g ){\bar \psi}_+ &=& 
 i\sqrt 2
\,\epsilon^+ {\bar {\cal D}}_{\zeta^-}{\bar \phi} 
-  2 Q \epsilon^- \sigma {\bar \phi}.     
\eea
It is understood that the supersymmetry transformation for 
the auxiliary field will be modified, and is given as:
\bea \label{sgF}
(\delta_s + \delta_g ) {\bar F} &=& 
- i\sqrt 2 \epsilon^- {\bar {\cal D}}_{\xi^-}{\bar \psi}_+ 
+ i\sqrt 2 \epsilon^+ {\bar {\cal D}}_{\zeta^-}{\bar \psi}_- \xx
&+& 2Q ( \epsilon^+ {\bar \psi}_+  {\bar \sigma} 
-  \epsilon^-  {\bar \psi}_- \sigma   ) 
- 2iQ {\bar \phi} ( \epsilon^+ \lambda_+ -  \epsilon^- \lambda_- ) \xx
&-& 2Q\,C^{00}\epsilon^+ {\bar {\cal D}}_{\xi^-}( 
{\bar \lambda}_+ {\bar \phi})  
- 2Q\,C^{11}\epsilon^-{\bar {\cal D}}_{\zeta^-}( 
{\bar \lambda}_- {\bar \phi})  \xx
&-& 2Q\,C^{10}\epsilon^- {\bar {\cal D}}_{\xi^-}( 
{\bar \lambda}_+ {\bar \phi})  
- 2Q\,C^{01}\epsilon^+ {\bar {\cal D}}_{\zeta^-}( 
{\bar \lambda}_- {\bar \phi}) .
\eea
Now, one can justify the choice of the terms given 
in eqn. (\ref{tildeF}). The first thing to
note is that, the only modification one expects for
the supersymmetry variation of $ {\bar F}$ is from 
additional terms added to  the definition of
the anti-chiral superfield which are proportional to
${\bar \theta}^-{\bar \theta}^+$. These are precisely
$C$-dependent terms given in eqn. (\ref{phiC}). 
Under supersymmetry variation, the terms in eqn. (\ref{phiC})
transform as:
\be \label{varphiC}
\delta{\bar \Phi_c} = 2\,{\bar\theta}^-{\bar\theta}^+
\left[\,C^{00}\,\epsilon^+ \partial_{\xi^-}
(\,{\bar \lambda}_+\,{\bar \phi})
\,+\,\,C^{11}\,\epsilon^- \partial_{\zeta^-}
({\bar \lambda}_-\,{\bar \phi})
\,+\,  C^{10} \epsilon^+\partial_{\zeta^-}
({\bar \lambda}_-\, {\bar \phi})
\,+\,  C^{01}\epsilon^- \partial_{\xi^-}
({\bar \lambda}_+\,{\bar \phi})\, \right].
\ee
From eqn. (\ref{sgF}), one can understand that the 
unique choice of terms in eqn. (\ref{tildeF}) is such that,
they add to the terms in eqn. (\ref{varphiC}) and form a 
gauge covariant derivative. This as we know is the ultimate aim
of writing supersymmetry transformations in the WZ gauge.
Thus, the choice of $C$-dependent terms made in eqn. (\ref{tildeF}) is
correct and unique.

Thus, eqn. (\ref{susySG}) and  (\ref{sgF}) summarize the 
${\cal N} =1/2$ supersymmetry transformations of the anti-chiral
multiplet and the corresponding transformations of the fields
in the chiral multiplet are given by  first four lines of 
eqn. (\ref{susy}). Now, as an explicit check, one can directly
calculate these supersymmetry transformations from eqn. (\ref{newphi})
and show that they are indeed correct.

These supersymmetry transformations will be used in section-4,
to check the invariance of the gauged linear sigma model action.

\section{Sigma Models with arbitrary  K\"ahler potential }                   

In our previous work~\cite{Chandrasekhar:2003uq}, we 
studied ${\cal N}=2$  supersymmetric
theories in two dimensions, characterized by an   
arbitrary  K\"ahler potential and superpotential with the 
superspace deformation as in   eqn. (\ref{Cdeformation}).
The discussion was limited to the case of a single chiral 
multiplet. It is interesting to generalize the discussion 
to include several multiplets. As this generalization leads 
to the construction of a sigma model and is also useful in
analyzing the target space geometry. 

Thus, in this section, 
we first generalize the results 
of~\cite{Chandrasekhar:2003uq} and study  sigma models 
characterized by an arbitrary  K\"ahler potential.
We show that the classical action admits a series expansion
in the determinant of the non(anti)commutativity parameter.
In fact, it is possible to write terms in this series expansion,
at an arbitrary order, in  a  closed form. 
In the
later part, we use a normal coordinate expansion to write 
the action in a covariant fashion.
In~\cite{Inami:2004sq}, a specific K\"{a}hler potential was considered,
and $CP^n$ models were analyzed in four dimensions.

\subsection{Expansion of the  K\"ahler potential}   

Let us start by giving the most general form of the 
classical action for supersymmetric 
sigma models on general K\"{a}hler manifolds:
\be \label{action}
I = \int d^2y \:d^4\theta \:
{\mathcal K}(\Phi^i,\bar{\Phi}^j),
\ee
where 
${\mathcal K}(\Phi^i,\bar{\Phi}^j)$ is the K\"{a}hler potential with
$\Phi^i$, $\bar{\Phi}^j$ denoting N chiral and anti-chiral superfields  
respectively.

To obtain the action in terms of
the component fields, the K\"{a}hler potential is Taylor 
expanded around the bosonic fields $\phi, {\bar \phi}$ as:
\bea \label{expand}
{\mathcal K}(\Phi,\bar{\Phi}) \!\!\!\!\!
&&={\mathcal K}(\phi^i,\bar{\phi^j}) 
+L^i \,{\mathcal K}^i\,+\,R^i \,{\mathcal K}_{,\bar j}
+ \frac{1}{2!}\,L^i*L^j\,{\mathcal K}_{,ij}
+ \frac{1}{2!}\,R^i*R^j
 {\mathcal K}_{,{\bar i}{\bar j}}
+ \frac{1}{2!} \left[L^i*R^j \right]
 {\mathcal K}_{,i{\bar j}} \xx
&&+\,\frac{1}{3!}
\,\left[\,L^i*L^j*R^k\,\right] \,
{\mathcal K}_{,ij{\bar k}} 
\,+\,  \frac{1}{3!}
\, \left[\,L^i*R^j*R^k\,\right] \,  
{\mathcal K}_{,i{\bar j}{\bar k}}
+\:\cdots\:+\,
\frac{1}{n!}\:L_*^{n}\: {\mathcal K}_{,i_1i_2\cdots i_n} \xx
&&+\frac{1}{m!}\:R_*^{m}\:{\mathcal K}_{,{\bar j}_1
{\bar j}_2\cdots {\bar j}_m}\:+\:\cdots\,
+\,\frac{1}{(n+m)!}\:\left[\,L_*^{n}*R_*^{m}\,\right] \:
{\mathcal K}_{,i_1i_2\cdots i_n{\bar j}_1
{\bar j}_2\cdots {\bar j}_m}
\:+\:\cdots\,.
\eea
A few remarks are in order,
regarding the expansion of the K\"{a}hler potential given above.
First, in eqn. (\ref{expand}),
$n,m$ are integers and we use the shorthand notation:
\be \label{kahler}
{\mathcal K}_{,i_1i_2\cdots i_n{\bar j}_1
{\bar j}_2\cdots {\bar j}_m} =
{ \frac{\partial^{(n+m)}{\mathcal K}}
{\partial \Phi^{i_1}\partial \Phi^{i_2}\cdots\partial \Phi^{i_n}
\partial {\bar \Phi}^{j_1}\partial {\bar \Phi}^{j_2}\cdots
\partial {\bar \Phi}^{j_m} } }|_{\Phi^i= \phi^i,\bar{\Phi}^i= {\bar \phi}^i},
\ee
for the derivatives of the K\"{a}hler potential with respect to the
chiral and anti-chiral superfields evaluated at 
$\Phi^i= \phi^i$ and $\bar{\Phi}^i= {\bar \phi}^i$. Note that the 
order of taking derivatives of the K\"{a}hler potential 
with respect to the chiral or anti-chiral superfields 
does not matter. In other words, 
${\mathcal K}_{,i_1i_2\cdots i_n{\bar j}_1
{\bar j}_2\cdots {\bar j}_m}$ is symmetric under any interchange
of $i$ indices or $j$ indices or an $i$ index with a $j$ index.
This symmetry will be useful while writing down
the action. Further, in eqn. \eq{expand}, the square brackets
$[\cdots]$ 
stand for all possible combinations of star product 
of $L^n$ with $R^m$, where 
$L_*^{n}~=~ L^{i_1}*L^{i_2}*......*L^{i_n}$  and
$R_*^{m}~=~ R^{i_1}*R^{i_2}*......*R^{i_m}$.
Explicitly\footnote{The notations used in~\cite{Chandrasekhar:2003uq}
to write down similar expressions are a bit different. One can use the
following coordinate changes to recover the results
in~\cite{Chandrasekhar:2003uq} : 
$A \ra \phi, {\bar A} \ra {\bar\phi} ; \bar{\psi}_L 
\ra  i\sqrt 2\psi_{-},\bar{\psi}_R 
\ra  i\sqrt 2\psi_{+} ; \psi_L \ra  i\sqrt 2\bar{\psi}_{-} ,  
\psi_R \ra  i\sqrt 2\bar{\psi}_{+} ; F \ra 2i F ,{\bar F} \ra  -2i {\bar F} $.
For the Grassmannian coordinates, the map is 
$\theta \ra - \theta^- , \chi \ra \theta^+, 
{\bar \theta} \ra -  {\bar \theta}^-, {\bar \chi} \ra {\bar \theta}^+ $. 
Further, one also has to take $ \partial_{\xi^-} \ra 2 \partial_{\xi^-} $ and 
$ \partial_{\zeta^-} \ra 2 \partial_{\zeta^-} $. 
},
\bea \label{L,R}
L^i = \Phi^i - \phi^i &=& + \sqrt 2 \theta^-\,\psi^i_- 
+ \sqrt 2 \theta^+\,\psi^i_+ - 2 \theta^-\,\theta^+\, F^i, \\
\label{Ronly}
R^i = \bar{\Phi}^i -{\bar \phi}^i &=&- \sqrt 2{\bar\theta}^-\,{\bar\psi}^i_- 
 - \sqrt 2{\bar\theta}^+\,{\bar\psi}^i_+  + 2i \theta^-{\bar\theta}^-
\partial_{\xi^-}{\bar \phi}^i + 2i\theta^+{\bar\theta}^+ 
\partial_{\zeta^-}{\bar \phi}^i \xx
&+& {\bar\theta}^-\,{\bar\theta}^+
(\,i2\sqrt 2\theta^+\,\partial_{\zeta^-}{\bar\psi}^i_-
-i2\sqrt 2\theta^-\,\partial_{\xi^-}{\bar\psi}^i_+ + 2\bar{F}^i +
4 \theta^-\,\theta^+\,\partial_{\xi^-}
\partial_{\zeta^-} {\bar \phi}^i  \,),
\eea
where we have suppressed the functional dependence of the component
fields on $(\xi^-,\zeta^-)$. The need for considering all possible
combinations (square brackets) in eqn. (\ref{expand}), has been
explained in great detail in~\cite{Chandrasekhar:2003uq}, and we
do not repeat it here. However, in the present case, there is an
additional permutational symmetry which we illustrate below.

Consider for instance, a term of the form $L^i*L^j$ in the expansion
of the K\"{a}hler potential in eqn. (\ref{expand}). If there was
only one chiral multiplet, this term would just be $L*L$. However,
if there are many chiral multiplets, then $L^i*L^j$ is not same
as $L^j*L^i$ due to the additional $C$-dependent terms coming 
from the star product. This can be seen by explicitly calculating
the two terms as shown below:
\bea \label{LiLj}
L^i*L^j &=&  -C^{00}\,\psi^i_- \psi^j_-  \,
-\,C^{11}\,\psi^i_+ \psi^j_+ 
- 2 \,(\theta^-\theta^+ - \frac{1}{2}C^{01})
\psi^i_-\psi^j_+ 
+ \,2 (\theta^-\theta^+ + \frac{1}{2} C^{10}) 
\psi^i_+\psi^j_-  \xx
&-& \sqrt 2(C^{00}\theta^+ + C^{01} \theta^- ) {\bar\psi}^i_- F^j  
+ \sqrt 2(C^{00}\theta^+ + C^{01} \theta^- ) {\bar\psi}^j_- F^i  
+ \sqrt 2(C^{10}\theta^+ + C^{11} \theta^- )  {\bar\psi}^i_+F^j \xx
&-& \sqrt 2(C^{10}\theta^+ + C^{11} \theta^- )  {\bar\psi}^j_+F^i
-  (\det \,C)\,F^i \, F^j,
\eea
and $L^j*L^i$ can be obtained by interchanging the indices $i$ and
$j$ in the above equation. 
Now, one can check that considering the permutation, 
$\left[\,L^i*L^j\,\right] = L^i*L^j+L^j*L^i$, there are lot of 
cancellations and only a few 
terms survive, as seen below:
\be \label{permute}
\left[\,L^i*L^j\,\right] 
= - 2 \left\{  2 \theta^-\theta^+
( \psi^i_- \psi^j_+ + \psi^j_- \psi^i_+ ) +
 (\det \,C)\,F^i \, F^j
\right\}.
\ee
Thus, it is
useful to repeat that,  apart from all possible combinations of
$L$'s and $R$'s considered in 
the expansion of the K\"{a}hler potential in 
eqn. (\ref{expand}), one has additional symmetry factors,
coming from the permutation of indices in either $L$'s or 
$R$'s. This will be discussed further, later on 
while writing down the action.

Hence, in what follows, we consider such permutations as in 
eqn. (\ref{permute}) to write down the action. First, using
the definitions of $L$ given in eqn. (\ref{L,R}), one can
generalize the result in eqn. (\ref{permute}) to  
calculate $\left[\,L^i*L^j*L^k\,\right]$. One can again 
show that there are many cancellations, by considering all 
possible permutations.
Proceeding in this
manner, one can check that the result in eqn. (\ref{permute}),
can be generalized to derive a general formula for the star
product of arbitrary number of $L$'s as 
shown below~\cite{Chandrasekhar:2003uq}:
\bea
\label{L2n}
[L_*^{2n}]&&=~  (-1)^n(\det~C)^{n-1} \,
\,\left[4n \theta^- \theta^+\{\,F^{i_1}F^{i_2}\cdots 
F^{i_{2n-2}}\:\psi^{i_{2n-1}}_-\,
\psi^{i_{2n}}_+ \,+\,{\rm perm.}\} \right.\xx
&&\left.~~~ + (2n)!\,  (\det~C)\,
F^{i_1}F^{i_2}\cdots F^{i_{2n}}  \, \right], \\
\label{L2n+1}
[L_*^{2n+1}]&&=~  (-1)^n(\det~C)^{n}\, \,
\left[2n\,\{ F^{i_1}F^{i_2}\cdots F^{i_{2n-1}}\,
\psi^{i_{2n}}_-\,\psi^{i_{2n+1}}_+ + {\rm perm.}\} \right.\xx
&&\left.~~~+ \{ F^{i_1}F^{i_2}\cdots F^{i_{2n}}\, 
L^{i_{2n+1}} + {\rm perm.}\}\right].
\eea
Note that in eqns. (\ref{L2n}), (\ref{L2n+1}) and in what
follows, the permutations are understood to be among the
$i_1\cdots i_n$ indices. In obtaining the identities given
in eqns. (\ref{L2n}) and (\ref{L2n+1}), we have also made use
of the fact that certain terms involving fermions are
anti-symmetric under the interchange of two indices where
as the derivatives of the K\"{a}hler potential are symmetric
under such interchange of indices.

For the star product of $R$'s, 
we derive the following results using eqn. (\ref{Ronly}):
\bea
\label{R2}
[R_*^2]&=&~ -4 {\bar\theta}^-\,{\bar\theta}^+\,
\Bigl[ 2{\bar\psi}^i_-{\bar\psi}^j_+ \,
-\,i\sqrt 2 \theta^+\,{\bar\psi}^i_-
  \,\partial_{\zeta^-} {\bar \phi}^j\,
  + i \sqrt 2\,\theta^-\,{\bar\psi}^i_+
\partial_{\xi^-} {\bar \phi}^j\,         
- 2\,\theta^-\,\theta^+\,\partial_{\xi^-} 
{\bar \phi}^i\partial_{\zeta^-} {\bar \phi}^j \xx
&+&\: {\rm perm.}\Bigr],  \\
\label{Rm}
R_*^{m}&=&~ 0,\quad {\rm for}~~ m \,>\, 2.
\eea

The other terms appearing in the expansion of the K\"{a}hler 
potential correspond to the star product of arbitrary powers 
of $L$'s and $R$'s. It is convenient to calculate the star
product of even and odd powers of $L$ with $R$ and $R_*^2$ 
separately. First we have:
\bea \label{LR}
[L^i\,*\,R^j]|_{{\bar\theta}^-\,
{\bar\theta}^+\,\theta^-\theta^+\,}&&= -8\,
\left( i\,\psi^i_-\,\partial_{\zeta^-}{\bar\psi}^j_-
\,+\, i\,\psi^i_+\,
\partial_{\xi^-}{\bar\psi}^j_+ \,+\, F^i\,{\bar F}^j\right), \\
\label{LRR}
\left[\,L^i*R^j*R^k\,\right]|_{{\bar\theta}^-\,
{\bar\theta}^+\,\theta^-\,\theta^+\,}
&&=~~\,-24\,i\,
\{ iF^i({\bar\psi}^j_-\,{\bar\psi}^k_+ + {\rm perm.})  \,
+\,\psi^i_-\,({\bar\psi}^j_-\,
\partial_{\zeta^-}{\bar \phi}^k+ {\rm perm.} )\, \xx
&&~~~~+\,\psi^i_+\,({\bar\psi}^j_+\,
\partial_{\xi^-}{\bar \phi}^k+ {\rm perm.})\,\}.
\eea
Note that in eqns. (\ref{LR}) and (\ref{LRR}), we have
only written the terms which are proportional to 
${\bar\theta}^-\,{\bar\theta}^+\,\theta^-\theta^+$, as 
only these terms contribute to the action, 
after integration over
the Grassmannian coordinates. Now, the identities
in eqns. (\ref{LR}) and (\ref{LRR})
can be generalized to\footnote{Note that, when we calculate 
terms of the kind $\left[\,L_*^{2n+1}\,*\,R_l\,\right]$ and
the ones to follow, we are actually writing down 
$\left[\,[L_*^{2n+1}]\,*\,R_l\,\right]$, where the additional
square bracket corresponds to permutations of indices of $L$.
However, in what follows we do not write this additional square bracket 
explicitly.}:
\bea
\label{L2n+1R}
&&\left[\,L_*^{2n+1}\,*\,R_l\,\right]|_{{\bar\theta}^-\,
{\bar\theta}^+\,\theta^-\,\theta^+\,} 
=~~\, \: 4(2n+2)(-1)^n\,(\det\,C)^{n}\,
\Bigl[\,2n\, \{ F^{i_1}F^{i_2}\cdots F^{i_{2n-1}}\,
\psi^{i_{2n}}_-\,\psi^{i_{2n+1}}_+\, \xx
&&~~+ {\rm perm.}\} 
\partial_{\xi^-}\partial_{\zeta^-}{\bar \phi}_l  
+\,i \{ F^{i_1}F^{i_2}\cdots F^{i_{2n}}
\psi^{i_{2n+1}}_-\,+ {\rm perm.}\}\,
\partial_{\zeta^-} {\bar \psi}_-\, \xx
&&~~+\,i \{ F^{i_1}F^{i_2}\cdots F^{i_{2n}}
\psi^{i_{2n+1}}_+\, 
+ {\rm perm.} \}
\partial_{\xi^-}{\bar \psi}_+ \,
+\,(2n+1)! F^{i_1}F^{i_2}\cdots F^{i_{2n+1}}
{\bar F}_l\,\Bigr], \\
\label{L2n+1RR}
&&\left[\,L_*^{2n+1}*R^j*R^k\,\right]|_{{\bar\theta}^-\,
{\bar\theta}^+\,\theta^-\,\theta^+\,} 
=~~ -4\,(2n+3)(2n+2)(-1)^n\,i\,(\det\,C)^{n}\,
\Bigl[\,-2\,n\,i( F^{i_1}F^{i_2}.. \xx
&&~~\times F^{i_{2n-1}}\psi^{i_{2n}}_-\,
\psi^{i_{2n+1}}_+\, 
+ {\rm perm.} )
(\partial_{\xi^-}{\bar \phi}^j\,
\partial_{\zeta^-}{\bar \phi}^k + {\rm perm.}) 
-\,i(   F^{i_1}F^{i_2}\cdots F^{i_{2n}}
\,\psi^{i_{2n-1}}_-\,+ {\rm perm.} ) \xx
&&~~\times({\bar\psi}^j_-\,i
\partial_{\zeta^-}{\bar \phi}^k + {\rm perm.}) 
- i(F^{i_1}F^{i_2}\cdots F^{i_{2n}} 
\psi^{i_{2n+1}}_+\,+ {\rm perm.})({\bar\psi}^j_+\,
\partial_{\xi^-}{\bar \phi}^k + {\rm perm.}) \xx
&&~~+\,(2n+1)!\,F^{i_1}F^{i_2}\cdots F^{i_{2n+1}}
({\bar\psi}^j_-\,{\bar\psi}^k_+\,+ {\rm perm.})\, \Bigr].
\eea
Similarly, the star product of even powers of $L$ with $R$ and
$R_*^2$ can be shown to be:
\bea
\label{L2nR}
&&\left[\,L_*^{2n}\,*\,R^k\,\right]|_{{\bar\theta}^-
\,{\bar\theta}^+\,\theta^-\,\theta^+\,} 
=~~\:4(2n+1)(-1)^n\,(\det\,C)^{n-1}\,
\Bigl[\, 2n \, {\bar F}^k
(F^{i_1}F^{i_2}\cdots F^{i_{2n-2}}
 \xx 
&\times&~~\psi^{i_{2n-1}}_-\,
\psi_{+,i_{2n}}\, + {\rm perm.}) +\,(2n)!(\det\,C)\,
F^{i_1}F^{i_2}\cdots F^{i_{2n}}
\partial_{\xi^-}\partial_{\zeta^-}{\bar \phi}^k \,\Bigr],\\
\label{L2nRR}
&&\left[\,L_*^{2n}*R^j*R^k\,\right]|_{{\bar\theta}^-\,
{\bar\theta}^+\,\theta^-\,\theta^+\,}
=\, -4(2n+2)(2n+1)(-1)^n
\left(\det~C\right)^{n-1}\,
\Bigl[\,2\,n (F^{i_1}F^{i_2} \xx
&&\times \cdots F^{i_{2n-2}} \psi^{i_{2n-1}}_-\,\psi_{+,i_{2n}}\, 
+ {\rm perm.})({\bar\psi}^j_-\,{\bar\psi}^k_+ + {\rm perm.})\,
- (2n)!\left(\det\,C\right)
F^{i_1}F^{i_2} \xx
&\times&~~~~~\cdots F^{i_{2n}}
(\partial_{\xi^-}{\bar \phi}^j\,
\partial_{\zeta^-}{\bar \phi}^k\,+ {\rm perm.}) \Bigr].
\eea
One can check that all the identities derived in 
eqns. (\ref{L2n})-(\ref{L2nRR}), go over to the ones 
derived in~\cite{Chandrasekhar:2003uq} for the case of 
a single chiral and antichiral supermultiplet, apart from
some permutational factors.

Now, substituting the results given in 
eqns. (\ref{L2n})-(\ref{L2nRR}) and performing integration over
the Grassmannian coordinates in the usual way,
it is possible to derive 
the full classical action for the ${\cal N}=2$ 
supersymmetric  
sigma model on a non(anti)commutative superspace. 

Before proceeding, we note that the sole effect of the 
permutations seen
in the identities in eqns. (\ref{L2n})-(\ref{L2nRR}), 
is to contribute an overall 
symmetry factor which cancels in the action. We 
illustrate this aspect for a couple of terms in the action 
and it will be clear that the argument can be generalized
to all the terms in the action. 
A term in the expansion of the K\"{a}hler 
potential (\ref{expand}) of the form, 
$\left[\,L^i*R^j*R^k\,\right]|_{{\bar\theta}^-\,
{\bar\theta}^+\,\theta^-\,\theta^+\,} {\mathcal K}_{,i\bar j \bar k}$, 
can be rewritten as  $\frac{1}{2!}\,
\left[\,L^i*[R^j*R^k]\,\right]|_{{\bar\theta}^-\,
{\bar\theta}^+\,\theta^-\,\theta^+\,} {\mathcal K}_{,i\bar j \bar k}$.
This can in turn be written as:
\be \label{sample}
\frac{1}{2!}\,24\,i\,\left[\
 F^i({\bar\psi}^j_-\,{\bar\psi}^k_+ + {\bar\psi}^k_-\,{\bar\psi}^j_+ )  \,
+\,\psi^i_-\,({\bar\psi}^j_-\,
\partial_{\zeta^-}{\bar \phi}^k+ {\bar\psi}^k_-\,
\partial_{\zeta^-}{\bar \phi}^j )\, 
+\,\psi^i_+\,({\bar\psi}^j_+\,
\partial_{\xi^-}{\bar \phi}^k+ {\bar\psi}^k_+\,
\partial_{\xi^-}{\bar \phi}^j )\,\right]
{\mathcal K}_{,i\bar j \bar k},
\ee
where in writing the above equation, 
we have used the result in eqn. (\ref{LRR}) and permuted terms
have been explicitly written down. However, 
as discussed before, one can use the symmetry of the   K\"{a}hler 
potential under the interchange of $j$ and $k$ indices, i.e.,
${\mathcal K}_{,i\bar j \bar k} = {\mathcal K}_{,i\bar k \bar j}$.
Using this symmetry, the result in eqn. (\ref{sample}) can be 
rewritten as: 
\be \label{sample1}
\:24\,i\,
\left[\,
F^i({\bar\psi}^j_-\,{\bar\psi}^k_+  )  \,
+\,\psi^i_-\,({\bar\psi}^j_-\,
\partial_{\zeta^-}{\bar \phi}^k )\, 
+\,\psi^i_+\,({\bar\psi}^j_+\,
\partial_{\xi^-}{\bar \phi}^k )\,\right]
{\mathcal K}_{,i \bar j \bar k}.
\ee
One can notice that the permutations in eqn. (\ref{sample}) contributed
an overall symmetry factor $2$ which canceled with $2!$ in the denominator
in  eqn. (\ref{sample1}). 
Since there are only two possible permutations of the terms of
the kind ${\bar\psi}^j_-\,{\bar\psi}^k_+$ etc., in eqn. (\ref{sample}), 
the symmetry factor one gets is $2$.

The above arguments can be easily generalized to get rid of the permutations
appearing in all the identities in eqns. (\ref{L2n})-(\ref{L2nRR}).
In fact, after using this symmetry of the K\"{a}hler potential,
the identities in eqns. (\ref{L2n})-(\ref{L2nRR}), will 
then go over to the ones derived 
in~\cite{Chandrasekhar:2003uq} for the case of 
a single chiral and antichiral supermultiplet, multiplied by 
appropriate overall symmetry factors.  

One can illustrate the above discussion by considering a more 
general term in the action. After using
the symmetry of the  K\"{a}hler potential as discussed above,
we have:
\bea
\label{L2n+1Rmodified}
&&\left[\,L_*^{2n+1}\,*\,R_l\,\right]|_{{\bar\theta}^-\,
{\bar\theta}^+\,\theta^-\,\theta^+\,} 
{\mathcal K}_{,i_1i_2\cdots i_{2n+1}\bar j } \xx
&&=\, 4\left(2n + 1 \right) !\:(2n+2)\,(-1)^n(\det\,C)^{n}\,
 F^{i_1}F^{i_2}\cdots F^{i_{2n-1}}\,
\Bigl[\,2n\,
\psi^{i_{2n}}_-\,\psi^{i_{2n+1}}_+\, 
\partial_{\xi^-}\partial_{\zeta^-}{\bar \phi}_l  \xx
&&+\,i F^{i_{2n}}
\psi^{i_{2n+1}}_-\,\,
\partial_{\zeta^-}{\bar \psi}_-\, 
+\, i F^{i_{2n}}
\psi^{i_{2n+1}}_+\, 
\partial_{\xi^-} {\bar \psi}_+ \, 
+\,F^{i_{2n}} F^{i_{2n+1}}{\bar F}_l\,
\Bigr]{\mathcal K}_{,i_1i_2\cdots i_{2n+1}\bar j },
\eea
where $\left(2n + 1 \right) !$ is the symmetry factor 
obtained after eliminating the permutations. However,
this symmetry factor will cancel after writing the
left hand side of eqn. (\ref{L2n+1Rmodified}) as
$\frac{1}{(2n+1)!}\left[\,[L_*^{2n+1}]\,*\,R_l\,\right]$.
This is nothing but the identity given in eqn. (\ref{L2n+1R}).
Similarly, various terms in the action can be rearranged
and the rest of the identities given in 
eqns. (\ref{L2n})-(\ref{L2nRR}), can be used in
an analogous fashion, while writing down the action. 

We further note that, in writing down the general identities 
given in eqns. (\ref{L2n})-(\ref{L2nRR}),
we have used the fact that for terms in the action
proportional to $\bar \theta^- \bar \theta^+ \theta^- \theta^+$, it
is possible to push all the $L$'s to one side and all
the $R$'s to the other side. The proof for the case
of a single chiral multiplet has been given 
in~\cite{Chandrasekhar:2003uq}, and can also be rigorously 
shown to be valid in the case of several chiral 
multiplet's as well.

\subsection{Classical Action in Normal Coordinates}   

Following the discussion in previous subsection and 
collecting all the results derived 
in eqns. (\ref{L2n})-(\ref{L2nRR}), and substituting them 
in the expansion 
of the  K\"{a}hler potential given in eqn. (\ref{expand}), 
we find that the action
can be divided into two parts as $I = I_0 + I_C$,
with $I_0$ and $I_C$ corresponding
to the $C$-independent and $C$-dependent parts respectively. 
First $I_0$ can be deduced to be:
\bea \label{I0}
I_0 &=& \int d^2x \,\Bigl[\,
(\frac{1}{2}\partial_{\xi^-}\phi^i
\partial_{\zeta^-}{\bar \phi}^j
+\frac{1}{2}\partial_{\zeta^-}\phi^i
\partial_{\xi^-}{\bar \phi}^j
+i\psi^i_-\partial_{\zeta^-}{\bar\psi}^j_-
+ i\psi^i_+\partial_{\xi^-}
{\bar\psi}^j_+ + F^i{\bar F}^j )
{\mathcal K}_{,i\bar j} + \psi^i_-\psi^j_+{\bar F}^k\xx
&\times& {\mathcal K}_{,ij\bar k} 
+ (i \psi^i_-{\bar\psi}^j_-\partial_{\zeta^-}{\bar \phi}^k
+ i \psi^i_+{\bar\psi}^j_+
\partial_{\xi^-}{\bar \phi}^k  
- F^i{\bar\psi}^j_-{\bar\psi}^k_+ )
{\mathcal K}_{,i\bar j \bar k}
+ (\psi^i_-{\bar\psi}^k_- 
\psi^j_+{\bar\psi}^l_+){\mathcal K}_{,ij\bar k \bar l}
\,\Bigr]\, ,
\eea 
where the derivatives of the K\"{a}hler potential are defined
in eqn. (\ref{kahler}). 
This action should be compared to the one in standard
literature~\cite{Braaten:is}.
$I_C$ can be derived in a similar fashion and is 
given as~\cite{Chandrasekhar:2003uq}:
\bea \label{IC}
I_C &=& -\sum_{n= 2}^{\infty}(-1)^n
(\det\,C)^{n-1}\int d^2x 
\frac{F^{i_1}F^{i_2}\cdots F^{i_{2n-2}} }{(2n-1)!}
\Bigl[\psi^{i_{2n-1}}_-
\psi^{i_{2n}}_+ {\bar F}^j 
 {\mathcal K}_{,i_1i_2\cdots i_{2n}\bar j }
+\,\psi^{i_{2n-1}}_-{\bar\psi}^j_-
\psi^{i_2n}_+{\bar\psi}^k_+  \xx
&\times&{\mathcal K}_{,i_1i_2\cdots i_{2n} \bar j \bar k}
\Bigr] 
+ \sum_{n=1}^{\infty}
(\det\,C)^n (-1)^n 
\int d^2x F^{i_1}F^{i_2}\cdots F^{i_{2n-1}}  
\Bigl[ \frac{1}{(2n)!} F^{i_{2n}} \partial_{\xi^-} 
\partial_{\zeta^-}{\bar \phi}^j
 {\mathcal K}_{,i_1i_2\cdots i_{2n}\bar j }   \xx
&\!\!\!\!+\!\!\!\!&\!\!\!\!\!\!\!\!
\frac {1}{(2n+1)!} 
\Bigl(\,2n\,\psi^{i_{2n}}_-
\psi^{i_{2n+1}}_+\partial_{\xi^-} 
\partial_{\zeta^-}{\bar \phi}^j 
+ i F^{i_{2n}}\psi^{i_{2n+1}}_-
\partial_{\zeta^-}{\bar\psi}^j_-  
+ iF^{i_{2n}}\psi^{i_{2n+1}}_+
\partial_{\xi^-} {\bar \psi}_+ 
+ F^{i_{2n}}F^{i_{2n+1}}{\bar F}^j\,\Bigr)  \xx
&\times&{\mathcal K}_{,i_1i_2\cdots i_{2n+1} \bar j}
+ \frac{1}{(2n)!}\,F^{i_{2n}} 
\partial_{\xi^-}{\bar \phi}^j\partial_{\zeta^-}{\bar \phi}^k\,
{\mathcal K}_{,i_1i_2\cdots i_{2n} \bar j \bar k}
+\frac{1}{(2n+1)!}\,
\Bigl(\,2n
\psi^{i_{2n}}_-
\psi^{i_{2n+1}}_+ \partial_{\xi^-}
{\bar \phi}^j \partial_{\zeta^-}{\bar \phi}^k \xx 
&-& F^{i_{2n}}F^{i_{2n+1}}{\bar\psi}^j_-{\bar\psi}^k_+ 
- F^{i_{2n}} {\bar\psi}^j_-\psi^{i_{2n+1}}_-
 \partial_{\zeta^-}{\bar \phi}^k 
- F^{i_{2n}} {\bar\psi}^j_+\psi^{i_{2n+1}}_+ 
\partial_{\xi^-}{\bar \phi}^k 
\,\Bigr) {\mathcal K}_{,i_1i_2\cdots i_{2n+1} 
\bar j \bar k}  \:\Bigr] \,.
\eea
The full action for the  
$ {\cal N}=2$ supersymmetric sigma model
on a non(anti)commutative superspace 
is thus given by  eqns. (\ref{I0}) and (\ref{IC}). Note that,
certain overall factors have been taken out in writing the
actions given above. Further, the actions given above, differ
from the ones in~\cite{Chandrasekhar:2003uq}, by some overall
factors and also by signs of some terms. One can explicitly
see the correspondence by using the map given in the footnote
above, between the variables
used in~\cite{Chandrasekhar:2003uq} and ones used here. 
Once these notational differences are taken into account, 
the action given by  eqns. (\ref{I0}) and (\ref{IC}) is same as the one 
given in~\cite{Chandrasekhar:2003uq}.

In~\cite{Chandrasekhar:2003uq}, the action has been shown to 
preserve the ${\cal N}=1/2$ supersymmetry of the theory,
in great detail. We do not repeat the calculations here. 
However, we have checked that the actions given in 
eqns. (\ref{I0}) and (\ref{IC}) are invariant under the 
${\cal N}=1/2$ supersymmetry transformations 
given in eqns. (\ref{susy}).

The power series expansion we see in the C-dependent part of
the action given in eqn. (\ref{IC}) is because of the
arbitrariness of the  K\"{a}hler potential. The
fact that the series can be summed and written in a closed form
is important. In fact, terms to an arbitrary order in $(\det\,C)$
can be easily deduced from  eqn. (\ref{IC}). However, at this
stage, it is not clear whether the full action can be written
in terms of covariant quantities. This will have to be taken
care while studying the quantum aspects of the theory. 
In other words, the question is, whether it is possible to 
see that the quantities
like  ${\mathcal K}_{,i_1i_2\cdots i_{2n} \bar j \bar k} $ etc.,
can be written in terms of proper geometric tensors. 

In the $C=0$ case, one can eliminate the auxiliary fields by 
their equations of motion and see that the action can be
written in terms of proper geometric quantities. 
To be precise, from eqn. (\ref{I0}), one can deduce that
$F^i = -\psi^{j}_- \psi^{k}_+ \, \Gamma^i_{jk} $ with 
similar relation for ${\bar F}$. Substituting these back
in eqn. (\ref{I0}), one arrives at:
\bea \label{I0G}
I_0 &=& \int d^2x \,\Bigl[\,(\partial_{\xi^-}\phi^i
\partial_{\zeta^-}{\bar \phi}^j
+i\psi^i_- {\bf D}_{\zeta^-}{\bar\psi}^j_-
+ i\psi^i_+ {\bf D}_{\xi^-}
{\bar\psi}^j_+)
\:g_{i\bar j} - \psi^i_-{\bar\psi}^k_- 
\psi^j_+{\bar\psi}^l_+  {\mathcal R}_{j\bar k i \bar l}
\,\Bigr].
\eea 
with covariant derivative defined as 
${\bf D}_{\xi^-} {\bar\psi}^i_+
= \partial_{\xi^-}{\bar\psi}^i_+ 
+  \Gamma^{\bar i}_{{\bar jk}} {\bar\psi}^j_+ \,
\partial_{\xi^-}{\bar \phi}^k $ and with a similar 
relation for the covariant derivative of 
${\bar\psi}^i_-$. Note that we have done a partial
integration in eqn. (\ref{I0G}). 
Further, we have used
the fact that, for  K\"{a}hler manifolds, the metric
can be obtained from the  K\"{a}hler potential as:
\be \label{metric}
g_{i\bar{j}} = \frac{\partial}{\partial \Phi^{i}} 
\frac{\partial}{\partial {\bar\Phi }^{j}} {\mathcal K}(\Phi,\bar{\Phi}).
\ee
Further, many components of the curvature tensor are zero. The
only non zero components are of the 
kind ${\mathcal R}_{j\bar k i \bar l}$ or 
 ${\mathcal R}_{\bar j k \bar i l}$.
There are many further simplifications and 
a brief collection of relavant formulae has been given in section-2.2
of~\cite{Chandrasekhar:2003uq}.

In the $C \neq 0$ case, it is not clear whether the auxiliary fields
can still be eliminated. The equation of motion of $F$ and ${\bar F}$,
which was found out for the $C=0$ case, may not be valid when the full
action in eqn. (\ref{IC}) is considered, because of infinite
number of terms in the action. 
Since, it is difficult to work with the $n^{th}$ order action,
below we first  analyze the action only to order  $(\det\,C)$. 
From eqn. (\ref{I0}) and (\ref{IC}), one can
write down the full action  $I = I_0 + I_C$, to order  $(\det\,C)$ as:
\bea \label{Ioc1}
I &=& \int d^2x \,\Bigl[\,(\frac{1}{2}\partial_{\xi^-}\phi^i
\partial_{\zeta^-}{\bar \phi}^j
+\frac{1}{2}\partial_{\zeta^-}\phi^i
\partial_{\xi^-}{\bar \phi}^j
+i\psi^i_-\partial_{\zeta^-}{\bar\psi}^j_-
+ i\psi^i_+\partial_{\xi^-}
{\bar\psi}^j_+ + F^i{\bar F}^j)
g_{i\bar j} + \psi^i_- \psi^j_+{\bar F}^k \xx
&\times&\Gamma_{ij\bar k} 
+ (i \psi^i_-{\bar\psi}^j_-\partial_{\zeta^-}{\bar \phi}^k
+ i \psi^i_+{\bar\psi}^j_+
\partial_{\xi^-}{\bar \phi}^k  
- F^i{\bar\psi}^j_-{\bar\psi}^k_+) \Gamma_{i\bar j \bar k}
+ (\psi^i_-{\bar\psi}^k_- 
\psi^j_+{\bar\psi}^l_+ ) \partial_i \Gamma_{j\bar k \bar l} 
+ (\det\,C)\:
F^{p} \xx
&\times& \Bigl[-\frac{F^{q}}{6}\Bigl\{
\psi^{l}_-
\psi^{m}_+ {\bar F}^j 
\partial_p  \partial_q \Gamma_{lm\bar j } 
+\psi^{l}_-{\bar\psi}^j_-
\psi^{m}_+{\bar\psi}^k_+ 
\partial_p  \partial_q \partial_l\Gamma_{m \bar j \bar k}
\Bigr\} -\frac{1}{2} F^{q} \partial_{\xi^-} 
\partial_{\zeta^-}{\bar \phi}^j\Gamma_{pq\bar j }  
-\frac {1}{6} 
\Bigl(2\psi^{q}_- \psi^{l}_+ \xx
&\times&
\partial_{\xi^-} 
\partial_{\zeta^-}{\bar \phi}^j 
+ i F^{q}\psi^{l}_-
\partial_{\zeta^-}{\bar\psi}^j_-  
+ iF^{q}\psi^{l}_+
\partial_{\xi^-}{\bar \psi}_+  
+ F^{q}F^{l}{\bar F}^j\,\Bigr) 
\partial_p \Gamma_{ql \bar j}         
- \frac{1}{2} \, F^{q} 
\partial_{\xi^-}{\bar \phi}^j\partial_{\zeta^-}{\bar \phi}^k\, 
\partial_p\Gamma_{q\bar j \bar k}  \xx
&-& \frac{1}{6}\,\Bigl(\,2 \psi^{q}_-
\psi^{l}_+ \partial_{\xi^-}
{\bar \phi}^j \partial_{\zeta^-}{\bar \phi}^k 
- F^{q}F^{l}{\bar\psi}^j_-{\bar\psi}^k_+ 
- F^{q} {\bar\psi}^j_-\psi^{l}_-
 \partial_{\zeta^-}{\bar \phi}^k 
- F^{q} {\bar\psi}^j_+\psi^{l}_+ 
\partial_{\xi^-}{\bar \phi}^k 
\,\Bigr) \partial_p  \partial_q \Gamma_{l 
\bar j \bar k}  \:\Bigr] \,,
\eea 
with the notation $\Gamma_{ij \bar k} = g_{l\bar k}\Gamma^l_{ij} $.
Further, the affine connections  
are obtained from the  K\"{a}hler
potential as: $\Gamma_{ij \bar k} = 
\frac{\partial}{\partial \Phi^{i}} 
\frac{\partial}{\partial \Phi^{j}} 
\frac{\partial}{\partial {\bar\Phi }^{k}} {\mathcal K} $. 

One can notice that various terms in the action in eqn. (\ref{Ioc1})
are non-covariant and the equation of motion of auxiliary fields
may be hard to find.
Thus, it is not possible to eleminate the auxiliary
fields even from the action (\ref{Ioc1}) to first order 
in  $(\det\,C)$, due to various non-covariant terms. 
In other words, the action (\ref{IC}) to all orders in  $(\det\,C)$ 
has to be expressed in terms of proper geometric quantities. 
Here, we show that it is possible to employ a normal coordinate expansion
and express the action in a manifestly covariant way.

In~\cite{Inami:2004sq}, a specific K\"{a}hler potential was
considered and the case of $CP^n$ models was discussed. After 
gauging the sigma model by introducing vector multiplets, 
it was possible to write the action in a closed form and 
the number of terms in the action was also finite. Hence, 
the auxiliary field equations of motion were simple to 
obtain. Thus, in section-4, we consider a simple K\"{a}hler
potential and show that the auxiliary fields appearing in the
chiral 2d superfields can be eleminated
by their equations of motion. However, we do not treat the
fields of the vector multiplet as auxiliary  and hence,
they are not eliminated, unlike the case in~\cite{Inami:2004sq}.

To ensure the covariance of various terms coming from the 
background field expansion of the sigma model action, 
Riemann normal coordinates were used 
in~\cite{Honerkamp:1971sh,Clark:ct}. This analysis 
was further used in the 
study of ultraviolet structure of the bosonic and supersymmetric
non-linear sigma models~\cite{Alvarez-Gaume:hn}. The need
for using normal coordinates was at the quantum level, in doing 
a background field expansion. 
Further, the discussion was 
explicitly for the case of ${\cal N} = 1$ sigma models. Although,
the method can be applied to ${\cal N} = 2$ sigma models, the results
were not manifestly covariant under ${\cal N} = 2$ supersymmetry.  
Recently, progress has been made
in this direction and 
the analysis has been generalized to  ${\cal N} = 2$ supersymmetric
sigma models on  K\"{a}hler manifolds~\cite{Higashijima:2000wz}. 
In~\cite{Higashijima:2000wz}, new  normal coordinates
were introduced and it was shown that the background field expansion
of the action can be written in a covariant manner. However, it was
pointed out that a manifestly supersymmetric expansion in these 
new coordinates is still not possible. The new coordinates are 
holomorphic and hence are, nevertheless, more suitable for 
K\"{a}hler manifolds.

Thus, below we show that, using the 
K\"{a}hler normal coordinates introduced 
in~\cite{Higashijima:2000wz}, the action can be written in terms
of covariant geometric quantities. Since, the origin of these
non-covariant terms is in the expansion of the K\"{a}hler 
potential, we try to identify the terms in this expansion
which give us the action  (\ref{Ioc1}).
Thus, we find that the sigma model action to first order in $(\det\,C)$
given in eqn. (\ref{Ioc1}), can be obtained from the following
terms in the expansion of the  K\"{a}hler potential:
\bea \label{expandc1}
I &=&  
\int d^2y \:d^4\theta 
\Bigl[\:
R^i \,{\mathcal K}_{,\bar i} 
\,+\,\frac{1}{2!}\,[\,R^i*R^j\,]\,{\mathcal K}_{,\bar i\bar j}
\,+\, \frac{1}{2!}
\,[\,L^i*R^j\,]\, {\mathcal K}_{,i\bar j}
+ \,\frac{1}{3!}
\,[\,L_*^2*R^k\,] \,  {\mathcal K}_{,ij\bar k} \xx
&+&\,  \frac{1}{3!}
\, [\,L^i*R_*^2\,] \,   {\mathcal K}_{,i\bar j\bar k}   
+ \,\frac{1}{4!}
\,[\,L_*^3*R^k\,] \,  {\mathcal K}_{,ijl\bar k}
\,+\, \frac{1}{4!}
\, [\,L_*^2*R_*^2\,] \,  {\mathcal K}_{,ik\bar j  \bar l}   
+ \,\frac{1}{5!}
\,[\,L_*^4*R^k\,] \,  {\mathcal K}_{,ijlm\bar k} \xx
&+&\, \frac{1}{5!}
\, [\,L_*^3*R_*^2\,] \,  {\mathcal K}_{,imk\bar j\bar l}   
\,+\, \frac{1}{6!}
\, [\,L_*^4*R_*^2\,] \,  {\mathcal K}_{,ikmp\bar j \bar l}   
\:\Bigr] + \: \cdots \:,
\eea
where ($\cdots$) in the above equation (\ref{expandc1}), 
corresponds to other higher order terms in $(\det\,C)$, 
in the expansion of the  K\"{a}hler potential.
Now, as we discussed, 
the sigma model action (\ref{Ioc1}), obtained from 
equation (\ref{expandc1}) above, is not invariant under 
general coordinate transformations. Thus, to write the
eqn. (\ref{expandc1}) in a covariant form, we rearrange the expansion
of the  K\"{a}hler potential and claim that the action can be
written as:
\be \label{expandnew}
I = \int d^2y \:d^4\theta 
\Bigl[
{\bar f}(\bar{\pi} ) +  g_{i{\bar j}}(\pi,{\bar \pi})\, \pi^i\,*\bar{\pi}^j
+  {\mathcal R}_{i\bar j k \bar l} \, \pi^i\,*\bar{\pi}^j*
 \pi^k\,*\bar{\pi}^l + \, \cdots \,\Bigr] ,
\ee
where we have introduced new superfields $\pi^i$ and $\bar{\pi}^j$,
which are given in terms of the old superfields as:
\bea \label{newsuper1}
\pi^i &=& L^i + \frac{1}{2!}\,L^l*L^m \, g^{i{\bar k}}\,
{\mathcal K}_{,lm{\bar k}}
+  \frac{1}{3!} L^l*L^m*L^n\, g^{i{\bar k}}\, 
{\mathcal K}_{,lmn{\bar k}} \xx
&+& \frac{1}{4!} L^l*L^m*L^n*L^p \, g^{i{\bar k}}\, 
{\mathcal K}_{,lmnp{\bar k}} + \: \cdots \: . \\
\label{newsuper2}
\bar{\pi}^j &=&  R^j +  \frac{1}{2!}\,R^k*R^l\, g^{m \bar j}\, 
{\mathcal K}_{,{\bar k}{\bar l} m} \,.
\eea
Note that, in terms of these new coordinates, one does not have
to consider various permutation and combination of indices. 
However, in new coordinates, one still continues to use the
star product as given in eqn. (\ref{star1}). As a consequence,
from eqn. (\ref{newsuper2}), one can show 
that the star product of more than three  ${\bar \pi}$'s vanishes.
Further, in 
eqn. (\ref{expandnew}), the function ${\bar f}$ is completely
anti-holomorphic and is given in terms of the old variables as
(first two terms on the right hand side of eqn.(\ref{expandc1}) ) :
\be \label{coordtr1}
{\bar f} = 
R^i \,{\mathcal K}_{,\bar i} 
\,+\,\frac{1}{2!}\,[\,R^i*R^j\,]\,{\mathcal K}_{,\bar i\bar j}\, .
\ee
To write the above function in terms of the new variables,
one has to invert the relations given in eqn. (\ref{newsuper2}) 
as shown:
\be
R^j = \bar{\pi}^j -  \frac{1}{2!}\,\bar{\pi}^k*\bar{\pi}^l
\, g^{m \bar j}\, {\mathcal K}_{,{\bar k}{\bar l} m}.
\ee
Now, using the above relations, the function ${\bar f}$ can be
written in terms of the new superfields. Generally speaking,
what one can actually do is to 
rearrange the expansion of the  K\"{a}hler potential given
in eqn. (\ref{expand}) as:
\be \label{expandnew2}
{\mathcal K}(\Phi,\bar{\Phi}) 
= {\mathcal K}(\pi,\bar{\pi}) + f(\pi) + 
{\bar f}(\bar{\pi} ) +  g_{i{\bar j}}\, \pi^i\,*\bar{\pi}^j
+  {\mathcal R}_{i\bar j k \bar l} \, \pi^i\,*\bar{\pi}^j*
 \pi^k\,*\bar{\pi}^l + \, \cdots \,,
\ee
where the functions the $f$ and ${\bar f}$ are holomorphic and 
anti-holomorphic respectively. The function ${\bar f}$ is
defined in eqn. (\ref{coordtr1}) above and
$f$ is given as: 
\bea \label{coordtr2}
 f(\pi) &=& L^i {\mathcal K}_{,i} 
+ \frac{1}{2!}\,[L^l*L^m] \, {\mathcal K}_{,lm}
+  \frac{1}{3!} [L^l*L^m*L^n]\,  
{\mathcal K}_{,lmn}
+ \frac{1}{4!} [L^l*L^m*L^n*L^p] \, 
{\mathcal K}_{,lmnp}   \, \xx
&+&\:\cdots \:\:+ 
[L_*^{n}]{\mathcal K}_{,i_1i_2\cdots i_{n}} 
+ \: \cdots .
\eea
Now, one can write the function $f$ in terms of new superfields
$\pi$, by inverting the relations given in eqn. (\ref{newsuper1}) as:
\be
L^i = {\pi}^i - \frac{1}{2!}\,{\pi}^l*{\pi}^m \, g^{i{\bar k}}\,
{\mathcal K}_{,lm{\bar k}} + \cdots \, ,
\ee
and using this relation in eqn. (\ref{coordtr2}). 
Notice that the functions $f$ and ${\bar f}$ explicitly contain 
many non-covariant quantities. However, these functions can be
absorbed into a redefinition of the   K\"{a}hler potential
by a  K\"{a}hler gauge transformation:
\be \label{kgauge}
{\mathcal K}'(\pi, {\bar \pi} ) 
= {\mathcal K}(\pi, {\bar \pi} ) +   f(\pi) + f(\bar{\pi}) . 
\ee
Now, one can check that the expansion of the  K\"{a}hler potential 
given in eqn. (\ref{expandnew}) generates all the terms in
eqn.  (\ref{expandc1}) to order  $(\det\,C)$. 
In addition  the term $ \pi^i\,*\bar{\pi}^j*
 \pi^k\,*\bar{\pi}^l,$ in eqn. (\ref{expandnew}) will
also give terms proportional to $(\det\,C)^2$ etc.,
The proof that an  expansion of the  kind 
given in eqn. (\ref{expandnew}) generates all the terms in the
action,
has been discussed in detail in~\cite{Higashijima:2000wz} and
can be checked in our case as well, by explicit calculation. 
The only difference compared to the case 
given in~\cite{Higashijima:2000wz}, is the presence of star
products instead of the ordinary product.
The expansion in new variables 
has the advantage that all the terms 
coming from it are covariant. This can be explicitly checked by
writing the transformations of various component fields under
holomorphic coordinate transformations and using the fact that
the coordinates (\ref{newsuper1}) transform as holomorphic tangent vectors on
target space~\cite{Higashijima:2000wz}.

It is important to note that 
the Riemann normal coordinates introduced 
in~\cite{Alvarez-Gaume:hn}, are inherently non-chiral. In other 
words, at a time, only one of the
old or new coordinates can be made chiral superfields. This
problem carries over to the case of  K\"{a}hler normal 
coordinates as well, although various quantities
are evaluated with respect to the bosonic
background. For instance, our old superfield $L$ is evaluated 
with respect to the bosonic background as it is given by
$\Phi - \phi$ and satisfies ${\bar D}_{\pm}\, L = 0$. However,
the expansion in new coordinates ($\pi,\bar{\pi} $) will not
preserve chirality as various geometric quantities will have
both holomorphic and anti-holomorphic indices~\cite{Higashijima:2000wz}.

The coordinate transformations
given in eqn. (\ref{coordtr1}) 
include all the terms in the expansion of the  K\"{a}hler potential.
Hence, the full $n^{\rm th}$ order action given in eqn. (\ref{IC})
can be written in a covariant manner. Further, 
it is possible to use the expansion of the  K\"{a}hler potential
in terms of the new variables given in eqn. (\ref{expandnew}) to
write the action in terms of the component fields. 
This action (\ref{IC}), written 
in terms of the new coordinates will be useful
while employing  background field methods to study 
the quantum structure of the theory. For this purpose,
one needs to calculate covariant expressions for the expansion of
various geometric quantities. Explicit expressions to a certain
order are given in~\cite{Higashijima:2000wz} for the $C=0$ case.
It should be interesting to find out similar expressions in our
case as well. Further, once the component form of the covariant
action is calculated from eqn. (\ref{expandnew}), it might be 
possible to find the equations of motion of the auxiliary 
fields $F$ and ${\bar F}$. 

Further, one can  
do a background field expansion for the simple case of 
a constant background, $\partial \phi_0 = 0$. 
This background field expansion of the action can be argued to 
be manifestly invariant under general holomorphic coordinate
transformations~\cite{Higashijima:2000wz}. However, in the 
present case, it may not
be manifestly invariant under ${\cal N}=1/2$ supersymmetry
transformations~\cite{Higashijima:2000wz}. In this section,
we have outlined how the ${\cal N} =2 $ sigma model action on
a non(anti)commutative superspace, can
be written in a covariant manner by transforming to the new
normal coordinates. It would be interesting to pursue these
issues further.

\section{Gauged Linear Sigma Models}   

In this section, we derive the classical action for Gauged linear
sigma models. We show that the action is invariant under
${\cal N}= 1/2$ supersymmetry transformations. 
The matter content of the theory is follows. We have $k$ chiral 
superfields $S^i$, which transform with charges $Q_{i}^a$ under the
$s$ vector multiplets $V^a$. As stated before, we only consider 
abelian gauge groups, which for our purposes will be $U(1)^s$. 

The superspace action corresponding to above multiplets can be written
as a Gauged linear Sigma model and consists 
of four parts~\cite{Witten:1993yc}:
\be
I = I_{kin} + I_{W} + I_{gauge} + I_{r,\theta}, 
\ee
where the terms are respectively, the kinetic term of the chiral
superfields, the superpotential interaction, the kinetic term of
the gauge fields, and the Fayet-Illiopoulos and theta terms. The
construction of all these terms is discussed below.
In writing the formulas, at some places 
we suppress the indices corresponding
to the number of multiplets for convenience.

\begin{flushleft}
{\underline {\ Chiral superfield Action}}
\end{flushleft}

The gauge invariant kinetic term for the 
Chiral superfields takes the form:
\be \label{kin}
I_{kin} = \int d^2y\,d^4\theta \:{\bar \Phi}*e^V*\Phi \, ,
\ee
where $d^2y = d\xi^- d\zeta^-$, 
$d^4\theta = d{\bar\theta}^- d{\bar\theta}^+ d\theta^- d\theta^+$ and 
the integrand can be evaluated owing to the results 
in eqn. (\ref{vpowers}), as shown below:
\be \label{Phipieces}
{\bar \Phi}*e^V*\Phi = {\bar \Phi}*\Phi +  {\bar \Phi}*V*\Phi + 
\frac{1}{2}{\bar \Phi}*V_*^2*\Phi \,.
\ee
We calculate each of the terms in eqn. (\ref{Phipieces}) 
separately and use them in the
action (\ref{kin}). The calculations are given in appendix B. Thus,
using the formulas derived in eqns. (\ref{ss})-(\ref{svvs}), 
in eqn. (\ref{kin}), and performing integration over the grassmannian
coordinates in the usual way, we find that the action can be 
written as:
\bea \label{chiralact}
I_{\mit kin}&=&\sum_i\int d\xi^- d\zeta^-
\left(\frac{1}{2} {\bar {\cal D}}_{\xi^-}\phi_i {\cal D}_{\zeta^-}{\bar \phi}_i 
+ \frac{1}{2} {\bar {\cal D}}_{\zeta^-}\phi_i {\cal D}_{\xi^-}{\bar \phi}_i 
+i{\bar \psi}_{-,i}{\cal D}_{\zeta^-}\psi_{-,i} 
+ i {\bar \psi}_{+,i} {\cal D}_{\xi^-} \psi_{+,i}+ F_i {\bar F}_i \right. \xx
&-& \left. 2\sum_a {\bar\sigma}_a \sigma_a Q_{i,a}{}^2 {\bar \phi}_i \phi_i
-  \sqrt 2\sum_aQ_{i,a}\left( {\bar\sigma}_a{\bar \psi}_{+i}\psi_{-i}
+\sigma_a{\bar \psi}_{-i}\psi_{+i}\right) + \sum_aD_a Q_{i,a}
{\bar \phi}_i\phi_i\right.\xx 
&-&\left. i\sqrt 2 \sum_a Q_{i,a}{\bar \phi}_i(\psi_{-,i}
\lambda_{+,a}-\psi_{+,i}\lambda_{-,a})
- i\sqrt 2 \sum_a Q_{i,a} \phi_i( {\bar \lambda}_{-,a}
{\bar \psi}_{+,i}- {\bar \lambda}_{+,a} {\bar \psi}_{-,i}) \right. \xx
&-& \left.  \sum_a Q_{i,a} \left( \sqrt 2 C^{00} ( i F \partial_{\xi^-}
\sigma_a {\bar \phi}_i  - {\bar \lambda}_{+,a}\psi_{-,i} 
{\bar {\cal D}}_{\xi^-} {\bar \phi}_i )  - \sqrt 2   C^{11}( - i F 
\partial_{\zeta^-}  {\bar \sigma}_a {\bar \phi}_i  \right. \right. \xx
&+& \left. \left. {\bar \lambda}_{-,a}\psi_{+,i} 
{\bar {\cal D}}_{\zeta^-} {\bar \phi}_i )  
- C^{01} ( i F_i {\bar \phi}_i \nu_{\xi\zeta} 
+ \sqrt 2 {\bar \lambda}_{+,a}\psi_{+,i}
{\bar {\cal D}}_{\xi^-}{\bar \phi}_i \right. \right. \xx
&-& \left. \left. \sqrt 2 {\bar \lambda}_{-,a}\psi_{-,i}
{\bar {\cal D}}_{\zeta^-}{\bar \phi}_i)  \right)
-  2\sum_aQ_{i,a}{}^2 (\det\,C) F_i{\bar \phi}_i  
{\bar \lambda}_{-,a} {\bar \lambda}_{+,a} \right)
\eea
where, ${\cal D}_{\xi^-} = \partial_{\xi^-} + \frac{i}{2}\nu_{\xi} ,$
${\cal D}_{\zeta^-} = \partial_{\zeta^-} + \frac{i}{2}\nu_{\zeta}$ are the 
gauge covariant derivatives and $ {\bar {\cal D}}_{\xi^-}, 
{\bar {\cal D}}_{\zeta^-}$ denote the corresponding complex 
conjugates respectively. The $C=0$ part of 
the kinetic action for the chiral 
superfields is seen
to be equivalent to the standard action given 
in~\cite{Witten:1993yc}. 
By using the transformation properties of the component fields, it can
be shown that the full kinetic action for the chiral superfields is
gauge invariant. 

Now, since the  $C=0$ part of action (\ref{chiralact}) is same 
as the one given in~\cite{Witten:1993yc}, one need not explicitly
show that this part is invariant under the ${\cal N}=1/2$ 
supersymmetry transformations. One can still check
this by using the supersymmetry transformations 
given in  eqn. (\ref{susySG}) and also the one
obtained by putting $C=0$ in eqn. (\ref{sgF}). 

For the case with $C \neq 0$, we know 
from eqn. (\ref{sgF}) and (\ref{lambdasusy}) that, only the 
supersymmetry variation 
of ${\bar F}$ and $\lambda_{\pm}$ have $C$-dependent terms. 
Thus, these terms are expected to cancel the variation of all
the $C$-dependent
terms of the action (\ref{chiralact}). This is exactly what we
show below.

Let us call the terms in the action depending on $C$ as $I_c$.
Then, the supersymmetry variation of these terms is:
\be \label{dIc}
\delta I_c =  (C^{00}\epsilon^+ + C^{10}\epsilon^- )\,\left( i
 {\bar \lambda}_+{\bar \lambda}_- + 
2{\bar {\cal D}}_{\xi^-}( 
{\bar \lambda}_+ {\bar \phi}    )  \right)
+ \,( C^{11}\epsilon^- + \,C^{01}\epsilon^+)\,\left( i
 {\bar \lambda}_+{\bar \lambda}_- + 2 {\bar {\cal D}}_{\zeta^-}( 
{\bar \lambda}_- {\bar \phi})  \right) .
\ee
Note that we have ignored the pieces whose
variations are trivially zero. For instance, the variation of
the term proportional to $(\det\,C)$ in the 
action (\ref{chiralact}) is zero identically. Now, one can
guess that the terms obtained in eqn. (\ref{dIc}), are exactly
canceled by the $C$-dependent coming from the following terms
of the action (\ref{chiralact}):
\be
- i\sqrt 2 {\bar \phi}\,\left( \psi_- \, ({\delta \lambda_+})\,  
-\, \psi_+\, ({\delta \lambda_-})\, \right) \:+\: F\, (\delta {\bar F}).
\ee
Using the supersymmetry transformations given in 
eqn. (\ref{sgF}) and (\ref{lambdasusy}), in the above 
equation, one can explicitly
show that the terms in  eqn. (\ref{dIc}) are exactly 
canceled. Thus, the chiral superfield action
(\ref{chiralact}), is invariant under the ${\cal N}=1/2$ 
supersymmetry of the theory.

\begin{flushleft}
{\underline { Gauge kinetic part}}
\end{flushleft}

A gauge invariant action for the vector superfields can be constructed
from the twisted chiral superfields as shown below~\cite{Witten:1993yc}:
\be
I_{gauge} = -\sum_a \frac{1}{4e_a^2}\, \int d^2y\,d^4\theta 
\: {\bar \Sigma}_a\,\Sigma_a \, ,
\ee
where $e_a$  $a= 1,\cdots,s$ are the gauge coupling constants, in
case one has several vector multiplets. Using 
the definitions of the twisted superfields given in 
eqns. (\ref{twistedC}) and (\ref{twistedAC}), the action can be
written in the component form as:
\bea \label{gauge}
I_{gauge} &=& - \sum_a \frac{1}{e_a^2} \int d^2y \,
\left( \frac{1}{2}\nu_{\xi\zeta,a}{}^2 
+ \frac{1}{2} D_a^2 + i  {\bar \lambda_{+,a}}
\partial_{\xi^-}\lambda_{+,a} 
+ i  {\bar \lambda_{-,a}}
\partial_{\zeta^-}\lambda_{-,a}  
\,-\,  \partial_{\xi^-}\sigma_a \partial_{\zeta^-}
{\bar \sigma}_{a}  \right. \xx
&-& \left. \frac{i}{2} \,C^{01}\,\nu_{\xi\zeta,a} \,{\bar \lambda_{+,a}}
\,{\bar \lambda}_{-,a} \:\right).
\eea
The kinetic energy for the gauge fields given above is, 
apart from some new C-dependent terms, same as the one given
in~\cite{Witten:1993yc}. The C-dependent term involving the
gauge field strength is gauge invariant on its own. It is useful
to compare this action with the dimensional reduction
of the action in~\cite{Seiberg:2003yz}.
Thus, the $C=0$ part can be taken to
be invariant under the ${\cal N}=1/2$ supersymmetry of theory.
This can as well be
explicitly checked by using the transformations given 
in eqn. (\ref{susySG}).

The $C\neq 0$ part of the action can also be shown to be 
${\cal N}=1/2$ supersymmetric as follows. 
We note
that, only the variations of $\lambda_{\pm}$ contain certain
C-dependent terms. Thus in eqn. (\ref{gauge}), 
the C-dependent terms obtained from $i  {\bar \lambda_{+,a}}
\partial_{\xi^-} (\delta \lambda)_{+,a} 
+ i  {\bar \lambda_{-,a}}
\partial_{\zeta^-} (\delta \lambda)_{-,a}  $ are:
\be
- {\bar \lambda_+}
\partial_{\xi^-}\left( ( C^{00} \epsilon^+ + C^{10} \epsilon^-)
 {\bar \lambda}_+{\bar \lambda}_- \right)
-   {\bar \lambda_-}
\partial_{\zeta^-}\left( ( C^{01} \epsilon^+ + C^{11} \epsilon^-)
 {\bar \lambda}_+{\bar \lambda}_-  \right).
\ee
One can see that the terms given in the above are identically
zero. Thus, one can guess that the  $C\neq 0$ part of the action
has to be supersymmetric on its own. It is straightforward
to check the $C\neq 0$ part of the action is invariant 
under ${\cal N}=1/2$ supersymmetry transformations as 
$\delta {\bar \lambda}_{\pm} = 0$ and other term also vanishes.
Thus, the gauge kinetic part of the action is invariant 
under the ${\cal N}=1/2$ supersymmetry of the theory. Further,
the gauge invariance of the action can also be explicitly
checked.

It should be interesting to take the $e^2 \ra \infty$ limit
where the fields of the vector multiplet become auxiliary and
can be eleminated by their equations of motion. The auxiliary
field equations of motion might have many $C$-dependent 
pieces~\cite{Inami:2004sq}, which might effect the target
space metric. In this manner, it would be possible to study
the consequences of various new $C$-dependent terms in the classical
action on the sigma model metric in $UV$ and 
$IR$ (see appendix B. of~\cite{Minwalla:2003hj}).

\begin{flushleft}
{\underline { $r$ and $\theta$ terms}}
\end{flushleft}
The Fayet-Iliopoulos (FI) and the theta angle terms can be obtained from
the twisted superfields as in~\cite{Witten:1993yc}. The FI term is the
vector superfield integrated over the whole of superspace. This term
should still be the same, since we
have not added any $C$-dependent term proportional 
to $\theta^2{\bar \theta}^2$ to the definition of vector
superfield given in eqn. (\ref{vmodified}). Thus, we have:
\be
I_{r,\theta} = -r_a\,\int d^2y\,  D^a \:+\:
\frac{\theta_a}{2\pi}\,\int d^2y\, \nu_{\xi\zeta,a} \:, 
\ee
where, as defined before,
$\nu_{\xi\zeta} = \partial_{\xi^-} \nu_{\zeta}\,
- \partial_{\zeta^-}\nu_{\xi} $.

\begin{flushleft}
{\underline { Superpotential terms}}
\end{flushleft}
If we assume an arbitrary superpotential, then the interaction
terms in the action turn out to have the form:
\be
I_{W} =  \int d^2x \:d^2\theta \: W(\Phi)
\,+\, \int d^2x\: d^2\bar{\theta}\: \bar{W}(\bar{\Phi}).
\ee
As was shown in~\cite{Chandrasekhar:2003uq}, the component form
of the superpotential can be obtained by expanding around the 
bosonic fields $\phi$ and $\bar \phi$ as:
\bea \label{W}
W(\Phi)|_{\theta^-\theta^+\,}
&=&  - \sum_{n=0}^{\infty}\:\frac{(-1)^n}{(2n+1)!}\: (\det\,C)^n
F^{i_1}F^{i_2}\cdots F^{i_{2n}}\: \left(\,
  F^{i_{2n+1}} \:W_{,i_1\cdots i_{2n+1}}\, \right. \xx
&+& \left. \, \psi^{i_{2n+1}}_- \psi^{i_{2n+2}}_+ 
\,W_{,i_1 \cdots i_{2n+2} } \, \right),
\eea
where as before, we use the notation $W_{,i} = 
 \frac{\partial W}{\partial \Phi^i}$ evaluated at $\Phi = \phi$.
In the above equations, we have only written down the 
terms proportional 
to  $\theta^+\theta^-$ and $\bar{\theta}^+ \,\bar{\theta}^-$
respectively. It is important to note that, the hermiticity of the
theory is spoiled due to the asymmetry of the holomorphic and the
anti-holomorphic parts of the superpotential~\cite{REY1}. This
can in fact be noted by looking at the asymmetric way in which
$F$ and $\bar F$ terms appear in the kinetic action.
Following the examples in four dimensions~\cite{REY1}, in the present 
case also, it may be possible consider supersymmetric
vacuua which come from $\bar{W}(\bar{\Phi})$ only, as $W(\Phi)$ may not
be stable due to radiative corrections. 

The anti-holomorphic part of the superpotential in the component
form is:
\bea
&&\bar{W}(\bar{\Phi})|_{\bar{\theta}^-
\,\bar{\theta}^+\,}
=  -\:{\bar F}^i\:  {\bar W}_{,i}
+ \,{\bar \psi}^i_- {\bar \psi}^j_+ \: {\bar W}_{,ij}
\eea
where $ {\bar W}_{,i} = \frac{\partial \bar{W}}
{\partial {\bar \Phi}^i}|_{\bar{\Phi}=\bar{\phi}}\, $.

It is possible to eliminate the auxiliary 
fields and write down
the $F$ term constraints as follow from their equations of motion:
\bea \label{eom}
F_i &=&  \,{\bar W}_{,i} , \xx
{\bar F}_i &=&   \sum_{n=0}^{\infty}\:\frac{(-1)^n}{(2n+1)!}\: (\det\,C)^n
F^{i_1}F^{i_2}\cdots F^{i_{2n-1}}\: \left(\, (2n+1)\,
 F^{i_{2n}} \: W_{,i_1 \cdots i_{2n+1}}\, \right. \xx 
&+& \left. \, (2n)\,\psi^{i_{2n+1}}_- \psi^{i_{2n+2}}_+ 
\,W_{,i_1 \cdots i_{2n+2} } \, \right) 
+ i \sum_a Q_{i,a} \left( \sqrt 2 C^{00} \partial_{\xi^-}
\sigma_a {\bar \phi}_i    + \sqrt 2   C^{11}  
\partial_{\zeta^-}  {\bar \sigma}_a {\bar \phi}_i  \right. \xx  
&-& \left. C^{01} {\bar \phi}_i \nu_{\xi\zeta} \right ) 
+  2\sum_aQ_{i,a}{}^2 (\det\,C) {\bar \phi}_i  
{\bar \lambda}_{-,a} {\bar \lambda}_{+,a}.
\eea
The $\bar F$ constraint contains new $C$-dependent pieces compared
to the standard case. In the present case, it was possible to
solve for the auxiliary field, unlike the case in the previous
section, where the K\"{a}hler potential was arbitrary.

We
Eliminate the auxiliary fields from the action and write
down the potential energy for the bosonic fields of the theory as:
\bea  \label{pot}
U &=&  \sum_{n=0}^{\infty}\:\frac{(-1)^n\,(\det\,C)^n}{(2n+1)!}\: 
  \: \left[\, (2n+1)\,(\,{\bar W}_{,i} )^{2n+1}
\: W_{,i_1\cdots i_{2n+1}}\, + \, (2n)\,
( \,{\bar W}_{,i} )^{2n}
\psi^{i_{2n+1}}_- \psi^{i_{2n+2}}_+    \right.    \xx
&\times& \left. \,W_{,i_1\cdots i_{2n+2} } \, \right] 
\:+\: \left[ \frac{D_a^2}{2e^2} + 2 \sigma_a {\bar \sigma}_a 
Q_{i,a}{}^2 \phi^i {\bar \phi}^i \right]. 
\eea
The potential seen above is again an expansion in powers of
$(\det\,C)$. Further, there are also many higher powers of the 
$\frac{\partial {\bar W}}{\partial {\bar \phi}^i}$. So, if 
${\bar W}(\bar \phi) $ is chosen to be zero, then no matter what
$W(\phi)$ is, the $F$ terms are zero and one is only left with
$D$ terms in the potential.

To draw more conclusions, let us look at the potential given in
eqn. (\ref{pot}) to first order in $(\det\,C)$:
\bea  \label{pot1}
U &=&  \frac{\partial {\bar W}}{\partial {\bar \phi}^i}
\frac{\partial W}{\partial \phi^i} -
(\det \,C)\frac{\partial {\bar W}}{\partial {\bar \phi}^i}\,
\frac{\partial {\bar W}}{\partial {\bar \phi}^j}\:
 (\,\frac{1}{2}\,\frac{\partial {\bar W}}{\partial {\bar \phi}^k}\: 
\frac{\partial^3 W}{\partial  \phi^i \partial\phi^j \partial\phi^k}\, 
+ \, \frac{1}{3}\,\psi^{k}_- \psi^{l}_+ 
\frac{\partial^4 W}{\partial  
\phi^i \partial\phi^j \partial\phi^k \partial\phi^l} \, ) \xx
&+&  \left( \frac{D_a^2}{2e^2} + 2 \sigma_a {\bar \sigma}_a 
Q_{i,a}^2 \phi^i {\bar \phi}^i \right). 
\eea
First, we see that the potential for the scalar fields also 
contains some fermionic pieces. In the $C=0$ case, these
fermionic pieces are absent and one can independently
look at the $F$-flatness and $D$-flatness conditions.
However, in the present case, the pieces depending on
$C$ come with a negative sign and hence, it is 
important to understand their role, while
looking for supersymmetric vacuua. In some simple cases, 
by suitable choice of $W$ and $\bar W$, the 
the fermionic pieces can be dropped.

Before proceeding, we note that, in eqn. (\ref{pot1}),
the $D$ term can be set to 
zero independently, as there are no other terms which depend on the 
gauge coupling:
\be
D = e^2\left( \,\sum Q\: \phi {\bar \phi} - r \:\right) = 0.
\ee
Notice that, the $D$ flatness condition is same as in the
$C=0$ theory. It is the analogue of this 
condition in~\cite{Witten:1993yc}, that gives the
the target space as $CP^{n-1}$. 

Now, one can make an appropriate choice for the superpotentials
and impose further conditions on the target space geometry.
For instance, following~\cite{Witten:1993yc}, one can take the matter
content to be, say, 
two chiral superfields $\Phi^1,\Phi^2$ of charge 1 each and one
chiral superfield $P$ of charge -2, such that the superpotential
$W(\Phi) = P*G(\Phi^1,\Phi^2) $ is gauge invariant,
quasi-homogeneous and satisfies
the constraints coming from $R$-symmetry invariance. Further, 
we can also choose ${\bar W}(\bar \Phi) = 
{\bar P}* {\bar G}({\bar\Phi}^1,{\bar\Phi}^2) $. In this case,
one can show that the fermionic terms drop out. 

Once again,
the $D$-term can be set equal to zero independently. The analogue
of this condition in terms of~\cite{Witten:1993yc} would give
the target space to be $CP^1$. Further, 
with the above choice of the superpotentials, evaluating 
eqn. (\ref{pot1}), one can have
new terms in the potential which depend on $(\det\,C)$.
For instance, the potential can have terms of the kind
$(\det\,C)\left( 
{\bar p}^2\,{\bar G}\,\frac{\partial G}{\partial \phi^1}
\frac{\partial G}{\partial \phi^2}
\right) $, apart from the standard terms which one normally
gets~\cite{Witten:1993yc}. Here, $\bar p,\phi^1,\phi^2 $
are the lowest components of corresponding superfields.
It should be interesting to 
vary the Fayet-Illiopoulos parameter $r$ and study the phases
of the above theory. In particular, to study the additional
restrictions put by the terms depending on $C$ on the target
space geometry.

\section{Discussion}                                    

To conclude, in 
this paper, we have extended the results of our previous 
work~\cite{Chandrasekhar:2003uq}, to write down the action 
for $D=2,{\cal N}=2$ sigma models
characterized by an arbitrary K\"{a}hler potential, 
on a non(anti)commutative
superspace, to include several chiral multiplets. 
Despite the fact that there are infinite number of 
terms, a general term in the action can be written down in a closed
form. This is
due to the fact that the action turns out to be
a series expansion in $(\det\,C) F$. 

It was shown that the 
action can be written in a manifestly covariant manner by
using the K\"{a}hler normal coordinates. This will be needed
while analyzing the quantum structure of the theory. It would
be interesting to apply the background field methods to study
the action using normal coordinates. Since, the  K\"{a}hler normal 
coordinates transform as holomorphic tangent vectors on the
target manifold, one expects that the background field expansion
will also be manifestly covariant. 
However, the background
field expansion may not preserve chirality and in this process 
invariance under the ${\cal N}=1/2$ supersymmetry transformations 
may also be lost. It is important to further study these
features, so as to address the question of renormalizability
of the theory.

In the second part, the analysis was extended to include 
Vector multiplets as well. We wrote down the classical
action for Gauged linear sigma models on non(anti)commutative
spaces. The gauge transformations and the supersymmetry 
transformations for the vector and chiral multiplets were
derived explicitly in the Wess-Zumino gauge. To ensure the
correctness of component calculations,
the action was  explicitly shown to be invariant under the 
${\cal N}=1/2$ supersymmetry transformations. The bosonic
potential of the thoery was shown to contain 
various higher powers of the derivatives of the superpotential.
The $D$-term constraint is still the same as in the $C=0$ theory. 
It would be interesting to 
turn on superpotentials considering various number of chiral 
multiplets and see what kinds of restrictions can be put on the target space
geometry. This would be the first step
to study the phases of the this model, 
in parallel to~\cite{Witten:1993yc}.

It is known~\cite{Seiberg:2003yz}, that supersymmetric theories
defined on non(anti)commutative superspace do not have a chiral
ring structure, due the absence of the ${\bar Q}$ supersymmetry.
This can also be inferred from the fact that product of an
arbitrary number of the chiral superfields does not vanish, in
general. This has some straightforward implications for topological
field theories. In the $C=0$ theory, it is know that if the 
left and (non-anamolous) right  $R$-symmetries are unbroken, then 
it is possible to have $A$ and $B$ twists. Many important 
properties of the untwisted models and several aspects of 
mirror symmetry have been studied from the
topological $A$ and $B$ models. For the present case, 
the absence of the chiral ring suggests that, 
it may not be possible to have the  standard $B$ twist. However,
it is possible to have the $A$ twist, where the operators are
in $Q = Q_+ + Q_-$ cohomology. It should be interesting to 
study these topological models.

There are other avenues one can explore. Taking 
the $e^2 \rightarrow \infty$ limit, one can look at the boundary 
terms generated from the gauged linear sigma model action. Since,
there are new $C$-dependent terms in the GLSM action, one expects 
new terms to be generated at the boundaries. These terms will 
play a crucial role while studying sigma models with boundaries
and hence, will be relevant in the study of $D$-brane using
GLSM's~\cite{Govindarajan:2000ef}.
On another front, one can look to solve the $D$-flatness
conditions and study the sigma model metric in the UV and IR.
Further, it should also be interesting to study
Closed string Tachyon condensation~\cite{Adams:2001sv}, in
this setting. 
We hope to come back to these issues in future.

\vskip 1.5cm

\begin{center}
{\bf Acknowledgments}
\end{center}
\vskip 0.3cm

I am thankful to Alok kumar for the suggesting the problem,
collaboration at the initial stages, for several useful 
discussions and comments on the manuscript. I am grateful to
the organizers of Spring School on Superstring theory
(13-25th March, 2004) at ASICTP, the HEP group 
of Institut de Physique, Universit\'{e} de Neuch\^{a}tel and
IACS, Kolkata for support and warm hospitality.
I would like to thank
M. Blau, J-P. Derendinger, S. Govindarajan, Biswanath
Layek, Avijit Mukherji, 
P. K. Panigrahi, Balram Rai, Koushik Ray, Siddharth Sen,
Soumitra Sengupta, Aninda Sinha and P. K. Tripathy
for valuable discussions. I would also like to thank N. Berkovits, 
Sunil Mukhi, Tapobrata Sarkar and S. Terashima for valuable
correspondences and references,
Aalok Misra for useful conversations and constant encouragement,
IOP String journal club members for discussions and the anonymous
referee for constructive suggestions.

\appendix

\sectiono{Some identities used in the text}


Some of the identities used in the text are given below. The can be 
derived using the definition of star product given 
in eqn. (\ref{star1}):
\bea \label{tt}
\theta^-*\theta^- &&=~ \frac{1}{2} C^{00},\\
\theta^+*\theta^+ &&=~ \frac{1}{2} C^{11}, \\
\label{tc}
\theta^-*\theta^+ &&=~ \theta^-\theta^+ - \frac{1}{2} C^{01}, \\
\theta^+*\theta^- &&=~ \theta^+\theta^- - \frac{1}{2} C^{10}, \\
\label{ttc}
\theta^-*\Bigl(\theta^-\theta^+\Bigr)    
&&=~ - \theta^-*\Bigl(\theta^+\theta^-\Bigr)~~~~~
~=~\frac{1}{2}\Bigl(\,C^{00} \theta^+ + C^{01}\theta^-\,\Bigr),\\
\label{cct}
\theta^+*\Bigl(\theta^+\theta^-\Bigr) &&=~
- \theta^+*\Bigl(\theta^-\theta^+\Bigr)~~~~~
~=~\frac{1}{2}\Bigl(\,C^{11} \theta^- + C^{10}\theta^+\,\Bigr),\\
\label{tctc}
\Bigl(\theta^-\theta^+\Bigr)*\Bigl(\theta^-\theta^+\Bigr) &&=~ 
- \Bigl(\theta^+\theta^-\Bigr)*\Bigl(\theta^-\theta^+\Bigr)~
~=~ -\frac{1}{4}\,(\det\,C).
\eea


\sectiono{Details of GLSM action}

Below we give some details of the calculation corresponding to the
chiral superfield action.
We write down the ${\bar\theta}^- {\bar\theta}^+ \theta^- \theta^+$ terms
coming from each of the pieces appearing on the right hand side
of eqn. (\ref{Phipieces}):
\bea \label{ss}
{\bar \Phi}*\Phi |_{{\bar\theta}^-
\,{\bar\theta}^+\,\theta^-\,\theta^+\,} 
&=& 4 \phi \partial_{\xi^-} \partial_{\zeta^-}{\bar \phi}
- 4i\,\psi_+\,\partial_{\xi^-}{\bar \psi}_+
\,-\, 4i\,\psi_-\,
\partial_{\zeta^-}{\bar \psi}_- \,-\, 4 F\,{\bar F} +
2i F\left[\sqrt 2 C^{00}  \right. \xx
&-& \left. \times \partial_{\xi^-}
(\sigma {\bar \phi})  
\sqrt 2 C^{11}\partial_{\zeta^-}(\bar \sigma \bar \phi)
+ C^{01}\partial_{\xi^-}(\nu_{\zeta}\bar \phi)
- C^{10}\partial_{\zeta^-}(\nu_{\xi}\bar \phi)\right] ,\\
\label{svs}
{\bar \Phi}*V*\Phi|_{{\bar\theta}^-
\,{\bar\theta}^+\,\theta^-\,\theta^+\,} &=& {\bar \phi}
\{ -  \phi (2D  +\,i\partial_{\zeta}\nu_{\xi} 
+\,i\partial_{\xi}\nu_{\zeta}) 
+ 2\sqrt 2 i \lambda_- \psi_+  -  2\sqrt 2 i \lambda_+ \psi_-  \xx
&-&  2\sqrt 2 i ( C^{01} {\bar \lambda}_+ \nu_{\xi} 
+  C^{11}{\bar \lambda}_- \nu_{\zeta} ) \psi_+ 
+  2\sqrt 2 i(  C^{00} {\bar \lambda}_+ \nu_{\xi} + 
 C^{10}{\bar \lambda}_- \nu_{\zeta} )\psi_- \} \xx
&+& \!\!\!\!\sqrt 2 {\bar \psi}_- \{ \sqrt 2  \nu_{\zeta}\psi_- 
+ 2 \sigma \psi_+  + 2i \phi  {\bar \lambda}_+ \}
+ \sqrt 2 {\bar \psi}_+ \{ \sqrt 2  \nu_{\xi}\psi_+ 
+ 2 {\bar \sigma} \psi_-  - 2i \phi  {\bar \lambda}_- \} \xx
&-& 2i \partial_{\xi^-}{\bar \phi} \{ \phi \nu_{\zeta} 
+ \nu_{\zeta} F C^{01} + \sqrt 2 \sigma F C^{00} 
- i \sqrt 2 {\bar \lambda}_+\psi_-  C^{00} 
+ i \sqrt 2  {\bar \lambda}_+\psi_+ C^{01} \} \xx
&-& 2i \partial_{\zeta^-}{\bar \phi} \{ \phi \nu_{\xi} 
- \nu_{\xi} F C^{10} - \sqrt 2 {\bar \sigma} F C^{11} 
+ i \sqrt 2 {\bar \lambda}_-\psi_+  C^{11} 
- i \sqrt 2  {\bar \lambda}_-\psi_- C^{10} \} \xx
\label{svvs}
{\bar \Phi}*V_*^2*\Phi |_{{\bar\theta}^-
\,{\bar\theta}^+\,\theta^-\,\theta^+\,}&=& 2 \phi {\bar \phi}(
-\nu_{\zeta}\nu_{\xi}+ 2 \sigma {\bar \sigma} ) 
+ 4 {\bar \phi}(\det C) {\bar \lambda}_- {\bar \lambda}_+ F.
\eea


\end{document}